\documentclass[aps,prd,twocolumn,superscriptaddress,showpacs]{revtex4}

\usepackage[dvips]{graphicx}
\usepackage{ulem}
\usepackage{float}
\usepackage[usenames]{color}
\usepackage{amsmath}
\usepackage{hyperref}
\usepackage[all]{hypcap}
\usepackage{float}

\begin{document}

\title{Strong Decays of observed $\Lambda_c$ Baryons in the $^3P_0$ Model}

\author{Jing-Jing Guo}

\affiliation{Department of Physics, Shanghai University, Shanghai 200444, China}

\author{Pei Yang}

\affiliation{Department of Physics, Shanghai University, Shanghai 200444, China}

\author{Ailin Zhang}
\email{zhangal@shu.edu.cn}

\affiliation{Department of Physics, Shanghai University, Shanghai 200444, China}

\begin{abstract}
All excited $\Lambda_c$ baryon candidates are systematically studied in a $^{3}P_{0}$ strong decay model. Possible Okubo-Zweig-Iizuka(OZI)-allowed strong decay channels of $\Lambda_c(2595)^+$, $\Lambda_c(2625)^+$, $\Lambda_c(2765)^+$ ($\Sigma_c(2765)^+$), $\Lambda_c(2860)^+$, $\Lambda_c(2880)^+$ and $\Lambda_c(2940)^+$ are given. The strong decay widths and some important branching ratios of these states are computed, and possible assignments of these $\Lambda_c$ baryons are given. (1), $\Lambda_c(2595)^+$ and $\Lambda_c(2625)^+$ are possibly the $1P$-wave charmed baryons $\Lambda_{c1}(\frac{1}{2}^-)$ and $\Lambda_{c1}(\frac{3}{2}^-)$, respectively. (2), $\Lambda_c(2765)^+$ ($\Sigma_c(2765)^+$) seems impossibly the $1P$-wave $\Lambda_{c}$, it could be the $2S$-wave or $1D$-wave charmed baryon. So far, the experimental information has not been sufficient for its identification. (3), $\Lambda_c(2860)^+$ seems impossibly $2S$-wave charmed baryon, it may be the $P$-wave $\tilde\Lambda_{c2}^{ }(\frac{3}{2}^-)$ or $\tilde\Lambda_{c2}^{ }(\frac{5}{2}^-)$, it could also be the $D$-wave $\check\Lambda_{c1}^{2}(\frac{1}{2}^+)$ or $\check\Lambda_{c1}^{2}(\frac{3}{2}^+)$. If the hypothesis that $\Lambda_c(2860)^+$ has $J^P={3\over 2}^+$ is true, $\Lambda_c(2860)^+$ is possibly the $D$-wave $\check\Lambda_{c1}^{2}(\frac{3}{2}^+)$ which has a predicted branching ratio $R=\Gamma(\Sigma_c(2520)\pi)/\Gamma(\Sigma_c(2455)\pi)=2.8$. (4), $\Lambda_c(2880)^+$ is impossibly a $1P$-wave or $2S$-wave charmed baryon, it may be a $D$-wave $\check\Lambda_{c3}^{2}(\frac{5}{2}^+)$ with $\Gamma_{total}=1.3$ MeV. The predicted branching ratio $R=\Gamma(\Sigma_c(2520)\pi)/\Gamma(\Sigma_c(2455)\pi)=0.35$, which is consistent with experiment. (5), $\Lambda_c(2940)^+$ is the $P$-wave $\tilde\Lambda_{c2}^{ }(\frac{3}{2}^-)$ or $\tilde\Lambda_{c2}^{ }(\frac{5}{2}^-)$, it is also possibly the $D$-wave $\check\Lambda_{c3}^{2}(\frac{5}{2}^+)$ or $\check\Lambda_{c3}^{2}(\frac{7}{2}^+)$. It is possible to distinguish the two assignments in $P$-wave or $D$-wave excitations through the measurement of $R=\Gamma(\Sigma_c(2520)\pi)/\Gamma(\Sigma_c(2455)\pi)$.
\end{abstract}

%\pacs{13.30.Eg, 14.20.Mr, 12.39.Jh}

\maketitle

\section{Introduction \label{sec:introduction}}
In the past years, in addition to established ground states, more and more highly excited charmed baryons have been observed by Belle, BABAR, CLEO and LHCb {\it et al}~\cite{pdg}. $\Lambda_c$ baryons have two light u, d quarks and one heavy $c$ quark inside. The two light quarks couple with isospin zero. The heavy quark symmetry works approximately in $\Lambda_c$ baryons, and the light quarks in $\Lambda_c$ baryons may correlate and make a diquark. The $\Lambda_c$ states provide an excellent window to explore the baryon structure and quark dynamics in baryons.

So far, in the review of particle physics\cite{pdg}, $\Lambda_c$, $\Lambda_c(2595)^+$, $\Lambda_c(2625)^+$, $\Lambda_c(2765)^+$
(or $\Sigma_c(2765)^+$), $\Lambda_c(2860)^+$, $\Lambda_c(2880)^+$, $\Lambda_c(2940)^+$ have been listed. The masses, total decay widths and possible decay channels of these $\Lambda_c$ are presented in Table.~\ref{table0}. The spins and parities of these $\Lambda_c$ states have not been measured by experiments. In order to identify these states, it is important to determine their $J^P$ quantum numbers and to learn their internal dynamics in every model.

\begin{table*}[htbp]
\setlength{\tabcolsep}{2.0mm}
\caption{Masses, decay widths (MeV), and possible strong decay channels of $\Lambda_c$.~\cite{pdg}}
\begin{tabular}{lcccccc} \hline \hline
& States  &$\mathbf{J}^P$&Mass &Width & Decay channels (experiment) & Decay channels in $^3P_0$ model. \\
\hline
&$\Lambda_c^+$       & $\frac{1}{2}^+$ &2286.46$\pm$0.14  & /          & weak  & /   \\
&$\Lambda_c(2595)^+$ & $\frac{1}{2}^-$ &2592.25$\pm$0.28   & 2.59$\pm$0.30$\pm$0.47& $\Sigma_c^{ ++,0}$$\pi^{-,+}$ &$\Sigma_c^{++,0}$$\pi^{-,+}$,$\Sigma_c^+$$\pi^0$    \\  %01
&$\Lambda_c(2625)^+$ &$\frac{3}{2}^-$  & 2628.11$\pm$0.19& $<$0.97                &$\Sigma_c^{++,0}$$\pi^{-,+}$ &$\Sigma_c^{++,0}$$\pi^{-,+}$,$\Sigma_c^+$$\pi^0$            \\  %02
&$\Lambda_c(2765)^+$ &$?^?$            &2766.6$\pm$2.4   &50           &/                        &$\Sigma_c^{(*)++,0,+}$$\pi^{-,+,0}$                                                            \\  %03
&$\Lambda_c(2860)^+$ &$\frac{3}{2}^+$  &$2856.1_{-1.7}^{+2.0}\pm0.5^{+1.1}_{-5.6}$            &$67.6_{-8.1}^{+10.1}\pm1.4^{+5.9}_{-20.0}$                   &$D^0p$        &$\Sigma_c^{(*)++,0,+}$$\pi^{-,+,0}$,$D^0p$,$D^+N$                          \\  %04
&$\Lambda_c(2880)^+$ &$\frac{5}{2}^+$  &2881.63$\pm$0.24 &$5.6_{-0.6}^{+0.8}$             &$\Sigma_c^{(*)++,0}$$\pi^{-,+}$,$D^0p$     &$\Sigma_c^{(*)++,0}$$\pi^{-,+}$,$D^0p$,$\Sigma_c^{(*)+}$$\pi^0$,$D^+N$     \\  %05
&$\Lambda_c(2940)^+$ &$?^?$            &$2939.6_{-1.5}^{+1.3}$  &$20_{-5}^{+6}$   & $\Sigma_c^{ ++,0}$$\pi^{-,+}$     &$\Sigma_c^{(*)++,0}$$\pi^{-,+}$,$D^0p$,$\Sigma_c^{(*)+}$$\pi^0$,$D^+N$     \\  %06
\hline\hline
\end{tabular}
\label{table0}
\end{table*}

Heavy baryons have been studied in many models, which could be found in some reviews~\cite{capstick,roberts,klempt,crede,cheng,hua} and references therein. Many tentative $J^P$ assignments to these $\Lambda_c$ states have been made in many models~\cite{capstick2,lutz,cheng2,he,zhu,zhu2,zhu3,zhu4,tolos,ebert,roberts,zhang0,chen,chen2,chen3,zhong,zhong2,dong1,dong2,dong3,liu,ortega,jianrong,oka,ping,npb926.(2018)467}. In addition to normal charmed baryon interpretations~\cite{capstick2,cheng2,zhu,zhu2,zhu3,zhu4,ebert,roberts,zhang0,chen,chen2,chen3,zhong,zhong2,oka,npb926.(2018)467}, there are also coupled-channel effect interpretations~\cite{lutz,tolos} and molecular state interpretations to these $\Lambda_c$~\cite{he,dong1,dong2,dong3,liu,ortega,jianrong,ping}. In Table~\ref{table1}, some possible $J^P$ assignments of $\Lambda_c$ within the charmed baryon interpretations are presented.

\begin{center}

\begin{table*}[t]
\caption{Some possible $J^P$ assignments of $\Lambda_c$.}
\begin{tabular}{p{1.0cm} p{2.5cm}*{8} {p{1.35cm}}}
\hline\hline
& Resonances & Ref.\cite{capstick2} & Ref.\cite{cheng2}& Refs.\cite{zhu,zhu2,zhu3,zhu4} & Ref.\cite{ebert} & Ref.\cite{roberts} & Refs.\cite{chen,chen2,chen3}& \cite{zhong,zhong2} & Ref.\cite{oka} \\
   \hline
   &$\Lambda_c $   & $\frac{1}{2}^+$ & $\frac{1}{2}^+$ & $\frac{1}{2}^+$    & $\frac{1}{2}^+$  & $\frac{1}{2}^+$ & $\frac{1}{2}^+$ &  $\frac{1}{2}^+$ &  $\frac{1}{2}^+$ \\
   &$\Lambda_c(2595)$ & $\frac{1}{2}^-$ & $\frac{1}{2}^-$ & $\frac{1}{2}^-$ & $\frac{1}{2}^-$  & $\frac{1}{2}^-$ & $\frac{1}{2}^-$ &  $\frac{1}{2}^-$ &  $\frac{1}{2}^-$\\
   &$\Lambda_c(2625)$ & $\frac{3}{2}^-$ & $\frac{3}{2}^-$ & $\frac{3}{2}^-$ & $\frac{3}{2}^-$  & $\frac{3}{2}^-$ & $\frac{3}{2}^-$ &  $\frac{3}{2}^-$ &  $\frac{3}{2}^-$ \\
   &$\Lambda_c(2765)$ ($\Sigma(2765)$) & $\frac{1}{2}^{+*}$ & $\frac{1}{2}^{+*}$ & $\cdots$   & $\frac{1}{2}^{+*}$   &  $\cdots$ & $\frac{1}{2}^{+*}$ &  $\cdots$  &  $\cdots$ \\
   &$\Lambda_c(2860)$ & $\cdots$   &   $\cdots$ & $\cdots$   & $\cdots$   &  $\cdots$  & $\frac{3}{2}^+$ &  $\cdots$ &  $\cdots$   \\
   &$\Lambda_c(2880)$ & $\cdots$   &   $\frac{5}{2}^+$  & $\frac{5}{2}^+$ & $\frac{5}{2}^+$ & $\cdots$ & $\frac{5}{2}^+$ &  $\frac{3}{2}^+$ &  $\frac{5}{2}^+$ \\
   &$\Lambda_c(2940)$ & $\cdots$   & ($\frac{3}{2}^+$, $\frac{5}{2}^-$) & $\cdots$   & $\frac{1}{2}^-$ &  $\cdots$ & $\cdots$  & $\frac{5}{2}^+$  &  $\frac{1}{2}^-,\frac{3}{2}^\pm,\frac{5}{2}^-$ \\
 \hline\hline
 \label{table1}
\end{tabular}
\end{table*}
\end{center}
For low-lying $\Lambda_c$, $\Lambda_c(2286)^+$ is believed the ground $1S$-wave charmed baryon with $J^P={1\over 2}^+$ without any doubt. $\Lambda_c(2595)^+$ and $\Lambda_c(2625)^+$ are popularly believed the $1P$-wave charmed baryon with $J^P={1\over 2}^-$ and $J^P={3\over 2}^-$, respectively. However, the $J^P$ assignments are different for highly excited $\Lambda_c(2765)^+$ ($\Sigma_c(2765)^+$), $\Lambda_c(2860)^+$, $\Lambda_c(2880)^+$ and $\Lambda_c(2940)^+$ in different literature though the $J^P={3\over 2}^+$ hypothesis is preferred for the $\Lambda_c(2860)^+$ and the $J^P={5\over 2}^+$ is constrained for the $\Lambda_c(2880)^+$ by LHCb collaboration~\cite{R.Aaij}. Furthermore, it is not clear yet whether $\Lambda_c(2765)^+$ ($\Sigma_c(2765)^+$) is an excited $\Lambda_c$ or $\Sigma_c$.

As known, a study of the strong decays of $\Lambda_c$ baryons is an important way to determine their $J^P$ quantum numbers. As a phenomenological method, the $^3P_0$ model was proposed to compute the OZI-allowed hadronic decay widths of hadrons~\cite{micu1969,yaouanc1,yaouanc2,yaouanc3}. There are also some attempts to make a bridge between the phenomenological $^3P_0$ model and QCD~\cite{swanson,ackleh,bonnaz}. The $^3P_0$ model has been employed to study the strong decays of $\Lambda_c$ baryons~\cite{zhu,zhang,zhang2,zhang3,zhang4}. In addition to the computation of strong decay widths, the dynamics and structure of the $\Lambda_c$ baryons have also been explored in these references. However, the studies aim at the separate analysis of one $\Lambda_c$ baryon or few observed $\Lambda_c$ baryons. The $\Lambda_c$ baryons have not been systematically analyzed in the $^3P_0$ model.

In this work, all the observed $\Lambda_c$ except for $\Lambda_c(2286)^+$ will be systematically examined as the $1P$-wave, $1D$-wave or $2S$-wave $\Lambda_c$ baryons from their strong decay properties in the $^3P_0$ model. In particular, their internal structure (especially the $\rho$-mode and $\lambda$-mode excitations ) will be paid attention to.

The paper is organized as follows. In Sec.II, the $^3P_0$ model is briefly introduced, some notations of heavy baryons and related parameters are indicated. We present our numerical results and analyses in Sec.III. In the last section, we give our conclusions and discussions.

\section{$^3P_0$ model, some notations and parameters\label{Sec: $^3P_0$ model}}
$^3P_0$ model is also known as a quark pair creation (QPC) model. It was first proposed by Micu\cite{micu1969} and further developed by Yaouanc {\it et al,}~\cite{yaouanc1,yaouanc2,yaouanc3}. The basic idea of this model assumes: a pair of $q\bar{q}$ are firstly created from the QCD vacuum
with quantum numbers $J^{PC}=0^{++}$; Subsequently, the created quark and antiquark recombine with the quarks from the initial hadron A to form two daughter hadrons B and C~\cite{micu1969}. The decays follow the OZI rule. For baryon decays, one quark of the initial baryon regroups with the created antiquark to form a meson, and the other two quarks regroup with the created quark to form a daughter baryon. There are three ways for the processes of recombination as follows,
\begin{eqnarray}
&A(q_1q_2q_3)+P(q_4\overline{q}_5)\to B(q_1q_4q_2)+C(q_3\overline{q}_5), \\
&A(q_1q_2q_3)+P(q_4\overline{q}_5)\to B(q_1q_4q_3)+C(q_2\overline{q}_5), \\
&A(q_1q_2q_3)+P(q_4\overline{q}_5)\to B(q_4q_2q_3)+C(q_1\overline{q}_5),
\end{eqnarray}
which are shown in Fig. 1, where each quark was numbered for a convenience.
\begin{figure*}
\begin{center}
\includegraphics[height=24cm,angle=0,width=16cm]{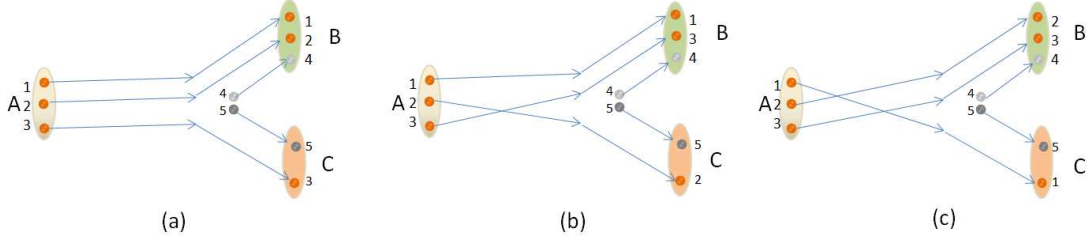}
\caption{Baryon decay process of $A\to B+C$ in the $^3P_0$ model. A is the initial
	baryon, B and C are the final baryon and meson, respectively.}
\end{center}
\end{figure*}
The two-body hadronic decay width $\Gamma$ for a baryon $A$ into $B$ and $C$ final states follows as
in the $^3P_0$ model~\cite{yaouanc3,zhang,zhang2,zhang3,zhang4,yang},
\begin{eqnarray}
\Gamma  &=& \pi ^2 \frac{|\vec{p}|}{m_A^2} \sum_{JL} |{\mathcal{M}^{JL}}|^2 \nonumber \\
&=&\pi ^2 \frac{|\vec{p}|}{m_A^2} \frac{1}{2J_A+1}\sum_{M_{J_A}M_{J_B}M_{J_C}} |{\mathcal{M}^{M_{J_A}M_{J_B}M_{J_C}}}|^2.
\end{eqnarray}
with $J=J_B+J_C$, $J_A=J_B+J_C+L$ and $M_{J_A}=M_{J_B}+M_{J_C}$. The partial wave amplitude $\mathcal{M}^{JL}$ is related to the helicity amplitude $\mathcal{M}^{M_{J_A}M_{J_B}M_{J_C}}$ via a Jacob-Wick formula~\cite{Jacob}. In the equation, $\vec{p}$ is the momentum of the daughter baryon in A's center of mass frame,
\begin{eqnarray}
 |\vec{p}|=\frac{{\sqrt {[m_A^2-(m_B-m_C )^2][m_A^2-(m_B+m_C)^2]}}}{{2m_A}},
\end{eqnarray}
$m_A$ and $J_A$ are the mass and total angular momentum of the initial baryon A, respectively. $m_B$ and $m_C$ are the masses of the final hadrons. The helicity amplitude $\mathcal{M}^{M_{J_A}M_{J_B}M_{J_C}}$ reads~\cite{zhu,zhang,zhang2,zhang4}
\begin{flalign}
 &\mathcal{M}^{M_{J_A } M_{J_B } M_{J_C }}\nonumber \\
 &=-F\gamma\sqrt {8E_A E_B E_C }  \sum_{M_{\rho_A}}\sum_{M_{L_A}}\sum_{M_{\rho_B}}\sum_{M_{L_B}} \sum_{M_{S_1},M_{S_3},M_{S_4},m}  \nonumber\\
 &\langle {J_{l_A} M_{J_{l_A} } S_3 M_{S_3 } }| {J_A M_{J_A } }\rangle \langle {L_{\rho_A} M_{L_{\rho_A} } L_{\lambda_A} M_{L_{\lambda_A} } }| {L_A M_{L_A } }\rangle \nonumber \\
 &\langle L_A M_{L_A } S_{12} M_{S_{12} }|J_{l_A} M_{J_{l_A} } \rangle \langle S_1 M_{S_1 } S_2 M_{S_2 }|S_{12} M_{S_{12} }\rangle \nonumber \\
 &\langle {J_{l_B} M_{J_{l_B} } S_3 M_{S_3 } }| {J_B M_{J_B } }\rangle \langle {L_{\rho_B} M_{L_{\rho_B} } L_{\lambda_B} M_{L_{\lambda_B} } }| {L_B M_{L_B } }\rangle \nonumber \\
 &\langle L_B M_{L_B } S_{14} M_{S_{14} }|J_{l_B} M_{J_{l_B} } \rangle \langle S_1 M_{S_1 } S_4 M_{S_4 }|S_{14} M_{S_{14} }\rangle \nonumber \\
 &\langle {1m;1 - m}|{00} \rangle \langle S_4 M_{S_4 } S_5 M_{S_5 }|1 -m \rangle \nonumber \\
 &\langle L_C M_{L_C } S_C M_{S_C}|J_C M_{J_C} \rangle \langle S_2 M_{S_2 } S_5 M_{S_5 }|S_C M_{S_C} \rangle \nonumber \\
&\times\langle\varphi _B^{1,4,3} \varphi _C^{2,5}|\varphi _A^{1,2,3}\varphi _0^{4,5} \rangle \times I_{M_{L_B } ,M_{L_C } }^{M_{L_A },m} (\vec{p}).
\end{flalign}
The conservation of the total angular momentum and the angular momentum of the light quarks freedom is indicated explicitly in the equation. $F$ is a factor equal to $2$ when each one of the two quarks in C has isospin ${1\over 2}$, and $F=1$ when one of the two quarks in C has isospin $0$.

In last equation, the matrix $\langle \varphi_B^{1,4,3} \varphi_C^{2,5}|\varphi_A^{1,2,3}\varphi_0^{4,5} \rangle$ of the flavor wave functions $\varphi_{i}$ $(i=A,B,C,0)$ can also be presented in terms of C-G coefficients of the isospin as follows~\cite{yaouanc3,zhang,yang}
\begin{flalign}
\scriptstyle
\langle\varphi_B^{1,4,3} \varphi_C^{2,5}|\varphi_A^{1,2,3}\varphi_0^{4,5} \rangle \scriptstyle=\mathcal{F}^{(I_A;I_BI_C)}\langle I_BI_B^3I_CI_C^3|I_AI_A^3\rangle
\end{flalign}
with
\begin{flalign}
\mathcal{F}^{(I_A;I_BI_C)} \nonumber
&= f \cdot (-1)^{I_{13}+I_C+I_A+I_2} \nonumber \\
&\times [\frac{1}{2}(2I_C  + 1)(2I_B  + 1)]^{1/2} \nonumber \\
&\times \begin{Bmatrix}
    {I_{13}} & {I_B} & {I_4}\\
    {I_C} & {I_2} & {I_A}\\ \end{Bmatrix}
\end{flalign}
where $f=(\frac{2}{3})^{1/2}$ for $u\bar u$ or $d\bar d$ created quark pair, and $f=-(\frac{1}{3})^{1/2}$ for $s\bar s$ created quark pair. $I_{A}$, $I_B$ and $I_C$ represent the isospins of the initial baryon, the final baryon and the final meson. $I_{12}$, $I_{3}$ and $I_{4}$ denote the isospins of relevant quarks. For example, the flavor matrix elements for $\Lambda_c^+\to\Sigma_c^{++,+,0}\pi^{-,0,+}$ and $\Lambda_c^+\to D^+n/D^0p$ are $\sqrt{1/6}$ and $\sqrt{1/3}$, respectively.

The space integral follows as
\begin{flalign}
I_{M_{L_B } ,M_{L_C } }^{M_{L_A } ,m} (\vec{p})&= \int d \vec{p}_1 d \vec{p}_2 d \vec{p}_3 d \vec{p}_4 d \vec{p}_5 \nonumber \\
&\times\delta ^3 (\vec{p}_1 + \vec{p}_2 + \vec{p}_3 -\vec{p}_A)\delta ^3 (\vec{p}_4+ \vec{p}_5)\nonumber \\
&\times \delta ^3 (\vec{p}_1 + \vec{p}_4 + \vec{p}_3 -\vec{p}_B )\delta ^3 (\vec{p}_2 + \vec{p}_5 -\vec{p}_C) \nonumber \\
& \times\Psi _{B}^* (\vec{p}_1, \vec{p}_4,\vec{p}_3)\Psi _{C}^* (\vec{p}_2 ,\vec{p}_5) \nonumber \\
& \times \Psi _{A} (\vec{p}_1 ,\vec{p}_2 ,\vec{p}_3)y _{1m}\left(\frac{\vec{p_4}-\vec{p}_5}{2}\right)
\end{flalign}

with a simple harmonic oscillator(SHO) wave functions for the baryons~\cite{capstick2,roberts2,zhang}
\begin{flalign}
\Psi_{}(\vec{p}_{})&=N\Psi_{n_{\rho} L_{\rho} M_{L_{\rho}}}(\vec{p}_{\rho}) \Psi_{n_{\lambda} L_{\lambda} M_{L_{\lambda}}}(\vec{p}_{\lambda}),
\end{flalign}
where $N$ represents a normalization coefficient of the total wave function. Explicitly,
\begin{flalign}
\Psi_{nLM_L}(\vec{p})&=\frac{(-1)^n(-i)^L}{\beta^{3/2}}\sqrt{\frac{2n!}{\Gamma(n+L+\frac{3}{2})}}\big(\frac{\vec{p}}{\beta}\big)^L \exp(-\frac{\vec{p}^2}{2\beta^2}) \nonumber\\
&\times L_n^{L+1/2}\big(\frac{\vec{p}^2}{\beta^2}\big)Y_{LM_L}(\Omega_p)
\end{flalign}
where $L_n^{L+1/2}\big(\frac{\vec{p}^2}{\beta^2}\big)$ denotes the Laguerre polynomial function, and $Y_{LM_L}(\Omega_p)$ is a spherical harmonic function. The relation between the solid harmonica polynomial $y _{LM}(\vec{p})$ and $Y_{LM_L}(\Omega_{\vec{p}})$ is $y _{LM}(\vec{p})=|\vec{p}|^L Y_{LM_L}(\Omega_p)$.

In order to describe three-body systems, a center of mass motion and a two-body systems of internal relative motion in the Jacobi coordinate~\cite{jacobi} are usually employed. As displayed in Fig. 2, $\vec{\rho}$ is the relative coordinate between the two light quarks (quark $1$ and $2$), and $\vec{\lambda}$ is the relative coordinate between the center of mass of the two light quarks and the charmed quark.
\begin{figure}[h]
\includegraphics[height=3cm,angle=0,width=7cm]{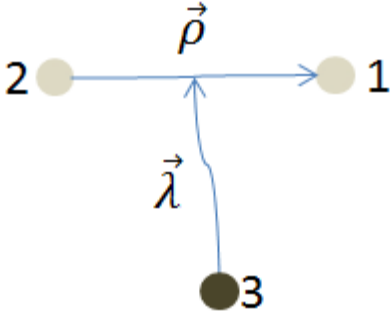}
\caption{Definitions of the Jacobi coordinates $\vec{\rho}$ and $\vec{\lambda}$. The quarks $1$ and $2$ are the light quarks, and quark $3$ is the heavy (charmed or bottomed) quark.}
\end{figure}

In the quark model with heavy quark symmetry~\cite{cheng2,zhu,zhu3,zhu4,zhang3,zhang4,oka,yang}, there are one $1S$-wave $\Lambda_{c}$, seven $P$-wave $\Lambda_{c}$, two $2S$-wave $\Lambda_{c}$, and seventeen $D$-wave $\Lambda_{c}$. The internal angular momentum of the $1S$-wave, $1P$-wave and $2S$-wave $\Lambda_{c}$ are presented in Table~\ref{table2}, where $\tilde\Lambda_{c0}^{\rho'}(\frac{1}{2}^+)$ and $\tilde\Lambda_{c0}^{\lambda'}(\frac{1}{2}^+)$ denote the radial excitation of a $\rho$-mode and a $\lambda$-mode, respectively.
The internal angular momentum of the $1D$-wave $\Lambda_{c}$ are presented in Table~\ref{table3}.

\begin{table}[ht]
\caption{Quantum numbers of $1S$-wave, $1P$-wave and $2S$-wave excited $\Lambda_c$.}
\begin{tabular}{ccc|cccccccc}
\hline\hline
&$N$  &~ Assignment ~ &$n_\rho$ &$n_\lambda$&~ $J$ ~& ~$J_l$ ~& ~$L_\rho$ ~&~ $L_\lambda$ ~&~ $L$  &~ $S_\rho$           \\
\hline
&1   &$\Lambda_{c0}(\frac{1}{2}^+)$    &0  &0    & $\frac{1}{2}$  & 0  &  0  &  0  &  0   &  0      \\
\hline
&2   &$\Lambda_{c1}^{ }(\frac{1}{2}^-)$        &0  &0    & $\frac{1}{2}$  & 1  &  0  &  1  &  1   &  0      \\
&3   &$\Lambda_{c1}^{ }(\frac{3}{2}^-)$        &0  &0    & $\frac{3}{2}$  & 1  &  0  &  1  &  1   &  0      \\
&4   &$\tilde\Lambda_{c0}^{ }(\frac{1}{2}^-)$  &0  &0 & $\frac{1}{2}$  & 0  &  1  &  0  &  1   &  1      \\
&5   &$\tilde\Lambda_{c1}^{ }(\frac{1}{2}^-)$  &0  &0 & $\frac{1}{2}$  & 1  &  1  &  0  &  1   &  1      \\
&6   &$\tilde\Lambda_{c1}^{ }(\frac{3}{2}^-)$  &0  &0 & $\frac{3}{2}$  & 1  &  1  &  0  &  1   &  1      \\
&7   &$\tilde\Lambda_{c2}^{ }(\frac{3}{2}^-)$  &0  &0 & $\frac{3}{2}$  & 2  &  1  &  0  &  1   &  1      \\
&8   &$\tilde\Lambda_{c2}^{1 }(\frac{5}{2}^-)$&0  &0& $\frac{5}{2}$  & 2  &  1  &  0  &  1   &  1      \\
\hline
&9   &$\tilde\Lambda_{c0}^{\rho'}(\frac{1}{2}^+)$  &1  &0 & $\frac{1}{2}$  & 0  &  0  &  0  &  0   &  0      \\
&10   &$\tilde\Lambda_{c0}^{\lambda'}(\frac{1}{2}^+)$  &0  &1 & $\frac{1}{2}$  & 0  &  0  &  0  &  0   &  0      \\
\hline\hline	
\label{table2}
\end{tabular}
\end{table}

\begin{center}
\begin{table}[htbp]
\caption{Quantum numbers of $1D$-wave excited $\Lambda_c$.}
\begin{tabular}{ccc|cccccccc} \hline \hline
&$N$	&~Assignment ~&$n_\rho$ &$n_\lambda$ &~$J$ ~&~$J_l$ ~ &~$L_\rho$~&~$L_\lambda$~&~$L$~&$S_\rho$       \\
\hline
&$1$	&$\Lambda_{c2}(\frac{3}{2}^+)$            &0  &0 &$\frac{3}{2}$ &2  &0  &2  &2  &0       \\  %01
&$2$	&$\Lambda_{c2}(\frac{5}{2}^+)$            &0  &0 &$\frac{5}{2}$ &2  &0  &2  &2  &0       \\  %02
&$3$	&$\hat\Lambda_{c2}(\frac{3}{2}^+)$        &0  &0 &$\frac{3}{2}$ &2  &2  &0  &2  &0       \\  %03
&$4$	&$\hat\Lambda_{c2}(\frac{5}{2}^+)$        &0  &0 &$\frac{5}{2}$ &2  &2  &0  &2  &0       \\  %04
&$5$	&$\check\Lambda_{c0}^{1}(\frac{1}{2}^+)$  &0  &0 &$\frac{1}{2}$ &0  &1  &1  &1  &1       \\  %05
&$6$	&$\check\Lambda_{c1}^{1}(\frac{1}{2}^+)$  &0  &0 &$\frac{1}{2}$ &1  &1  &1  &1  &1       \\  %06
&$7$	&$\check\Lambda_{c1}^{1}(\frac{3}{2}^+)$  &0  &0 &$\frac{3}{2}$ &1  &1  &1  &1  &1       \\  %07
&$8$	&$\check\Lambda_{c2}^{1}(\frac{3}{2}^+)$  &0  &0 &$\frac{3}{2}$ &2  &1  &1  &1  &1       \\  %08
&$9$	&$\check\Lambda_{c2}^{1}(\frac{5}{2}^+)$  &0  &0 &$\frac{5}{2}$ &2  &1  &1  &1  &1       \\  %09
&$10$	&$\check\Lambda_{c1}^{0}(\frac{1}{2}^+)$  &0  &0 &$\frac{1}{2}$ &1  &1  &1  &0  &1       \\  %10
&$11$	&$\check\Lambda_{c1}^{0}(\frac{3}{2}^+)$  &0  &0 &$\frac{3}{2}$ &1  &1  &1  &0  &1       \\  %11
&$12$	&$\check\Lambda_{c1}^{2}(\frac{1}{2}^+)$  &0  &0 &$\frac{1}{2}$ &1  &1  &1  &2  &1       \\  %12
&$13$	&$\check\Lambda_{c1}^{2}(\frac{3}{2}^+)$  &0  &0 &$\frac{3}{2}$ &1  &1  &1  &2  &1       \\  %13
&$14$	&$\check\Lambda_{c2}^{2}(\frac{3}{2}^+)$  &0  &0 &$\frac{3}{2}$ &2  &1  &1  &2  &1       \\  %14
&$15$	&$\check\Lambda_{c2}^{2}(\frac{5}{2}^+)$  &0  &0 &$\frac{5}{2}$ &2  &1  &1  &2  &1       \\  %15
&$16$	&$\check\Lambda_{c3}^{2}(\frac{5}{2}^+)$  &0  &0&$\frac{5}{2}$  &3  &1  &1  &2  &1       \\  %16
&$17$	&$\check\Lambda_{c3}^{2}(\frac{7}{2}^+)$  &0  &0 &$\frac{7}{2}$ &3  &1  &1  &2  &1       \\  %17
\hline\hline
\end{tabular}
\label{table3}
\end{table}
\end{center}

In these tables, $L_\rho$ denotes the orbital angular momentum between the two light quarks, $L_\lambda$ denotes the orbital angular momentum between the charm quark and the two light quark system, $S_\rho$ denotes the total spin of the two light quarks. $L$ is the total orbital angular momentum of $L_\rho$ and $L_\lambda$ ($L$ =$L_\rho$ + $L_\lambda$), and $J_l$ is the total angular momentum of $L$ and $S_\rho$ ($J_l$ = $L$ + $S_\rho$). $J$ is the total angular momentum of the baryons ($J$ = $J_l$ + ${\frac{1}{2}}$). In $\tilde\Lambda_{cJ_l}^{\ L}$($\tilde\Sigma_{cJ_l}^{\ L}$), a superscript $\ L$ denotes the total angular orbital momentum, a tilde indicates $L_\rho=1$, and the one without a tilde indicates $L_\rho=0$. More details about the notations could be found in Refs.~\cite{roberts2,zhu,zhang3,zhang4}
	
In the $^3P_0$ model, the $q\bar{q}$ quark pair created from the vacuum may be $u\bar{u}$, $d\bar{d}$ or $s\bar{s}$. So far, there is no sign of an $s\bar{s}$ creation in observed strong decay channels of $\Lambda_c$ states. In addition to masses, decay widths, experimentally observed strong decay channels, theoretically predicted strong decay channels of all the $\Lambda_c$ states are also given in Table~\ref{table0}. Masses of relevant mesons and baryons involved in our calculation are presented in Table~\ref{table4}~\cite{pdg}.

\begin{table}[H]
\caption{Masses of mesons and baryons involved in the decays~\cite{pdg}}
\begin{tabular}{p{0.0cm} p{2.0cm}p{2.0cm}|p{2.0cm}p{2.0cm}}
\hline\hline
		&State              &Mass (MeV)  & State          &Mass (MeV)\\
		\hline
		&$\pi^{\pm}      $ &139.570     &$\Sigma_c(2520)^{++}$    & 2518.41  \\
		&$\pi^{0}        $ &134.977     &$\Sigma_c(2520)^{+}$     & 2517.5   \\
		&$K^{\pm}        $ &493.677     & $\Sigma_c(2520)^{0}$    & 2518.48  \\
		&$K^{0}          $ &497.611     & $\Sigma_c(2455)^{++}$   &2453.97  \\
		&$\Lambda_c^{+}$ & 2286.46      & $\Sigma_c(2455)^{+}$    &2452.9  \\
		&$D^{0}$         & 1864.84      &$\Sigma_c(2455)^{0} $    &2453.75  \\
		& $D^{+} $       & 1869.59      &$\Sigma_c(2765)^{++}$    &2766.6 \\
		&$\Sigma_c(2765)^{+} $   &2766.6&$\Sigma_c(2765)^{0} $   &2766.6\\
		&$\Sigma_c(2800)^{+} $   &2792  &-                        &-            \\
\hline\hline
\end{tabular}
\label{table4}
\end{table}
		
The parameters are chosen as follows. The dimensionless pair-creation strength $\gamma=13.4$. The $\beta_{\lambda,\rho}=600$ MeV in the $1S$-wave baryon wave functions are chosen,  the $\beta_{\lambda,\rho}=500$ MeV in the $P$-wave baryon wave functions are chosen, and the $\beta_{\lambda,\rho}=400$ MeV in the $2S$-wave and $D$-wave baryon wave functions are chosen. These $\beta_{\lambda,\rho}$ are consistent with those in Refs.~\cite{godfrey,zhu,godfrey2,godfrey3,zhang}. The $R=2.5$ GeV$^{-1}$ in the harmonic oscillator wave functions of $\pi/K$ meson and $R=1.67$ GeV$^{-1}$ for $D$ meson~\cite{godfrey,zhu,godfrey2,godfrey3,zhang}.

\section{Strong decays of $\Lambda_c$\label{Sec: numerical results}}
\subsection{$\Lambda_c(2595)$ and $\Lambda_c(2625)$}

$\Lambda_c(2595)^+$ and $\Lambda_c(2625)^+$ were first discovered by the ARGUS Collaboration at the $e^+$$e^-$ storage ring DORIS II at DESY~\cite{Albrecht:1993pt}, and subsequently confirmed by E687~\cite{Frabetti:1993hg} and CLEO~\cite{Edwards:1994ar} Collaborations.

$\Lambda_c^+\pi\pi$ and its submode $\Sigma_c(2455)\pi$ are the only allowed strong decays of $\Lambda_c(2595)^+$. $\Lambda_c^+\pi\pi$ results from a two steps process $\Lambda_c(2595)\to\Sigma_c(2455)\pi$ with $\Sigma_c(2455)\to\Lambda_c\pi$, and a direct $\Lambda_c^+\pi\pi$ three-body decay with fraction about $18\pm10\%$. The branching fractions $\Gamma(\Lambda_{c}(2595)^{+} \to \Sigma_c^{++} \pi^{-})/\Gamma_{total}=24\pm7\%$ and $\Gamma(\Lambda_{c}(2595)^{+} \to \Sigma_c^{0} \pi^{+})= 24\pm7\%$~\cite{pdg}.

$\Lambda_c^+\pi\pi$ and its submode $\Sigma_c(2455)\pi$ are also the only allowed strong decays of $\Lambda_c(2625)^+$. In contrast to $\Lambda_c(2595)^+$, the branching fraction of the direct three-body decay mode $\Lambda_c^+\pi\pi$ of $\Lambda_c(2625)^+$ is large, while the branching fraction $\Gamma(\Lambda_c(2625)^+ \to \Sigma_c^{++}\pi^- )/ \Gamma_{total}$ or $\Gamma(\Lambda_c(2625)^+ \to \Sigma_c^{0}\pi^+ )/ \Gamma_{total}$ is less than $5\%$~\cite{pdg}, which means that the decay width $\Gamma(\Lambda_c(2625)^+ \to \Sigma_c^{++}\pi^- )$ or $\Gamma(\Lambda_c(2625)^+ \to \Sigma_c^{0}\pi^+ )$ is less than $0.05$ MeV.

$\Lambda_c(2595)^+$ and $\Lambda_c(2625)^+$ are believed the low-lying $P$-wave $\Lambda_c$, and form a doublet $\Lambda_{c1}(\frac{1}{2}^-,\frac{3}{2}^-)$~\cite{cheng2,ebert,zhu2}. Their $J^P$ are supposed ${1\over2}^-$ and ${3\over2}^-$, respectively~\cite{pdg}. In our analyses, all the hypothesises that $\Lambda_c(2595)^+$ and $\Lambda_c(2625)^+$ are the low-lying $1P$-wave, $2S$-wave, and $1D$-wave charmed baryons are examined. In Table~\ref{table5}, the numerical results of the decay widths of $\Lambda_c(2595)^+$ as the $1P$-wave and $2S$-wave states are given. Similar numerical results for $\Lambda_c(2625)^+$ are presented in Table~\ref{table6}. In Table~\ref{table7} and Table~\ref{table8}, the numerical results of the decay widths of $\Lambda_c(2595)^+$ and $\Lambda_c(2625)^+$ as $D$-wave charmed baryons are given, respectively. In these tables, some branching ratios are also given.

\begin{center}
	\begin{table*}[t]
		\caption{Decay widths (MeV) of $\Lambda_c(2595)^+$ as $1P$-wave and $2S$-wave charmed baryons. $\mathcal{B}=\Gamma(\Sigma_c^{++}\pi^{-})/\Gamma_{total}$. }
		\begin{tabular}{lccccccc}
			\hline\hline
			&N~&$\Lambda_{cJ_l}(J^P)$ &$\Sigma_c^{++}\pi^- $ &$\Sigma_c^{0}\pi^+ $ &$\Sigma_c^{+}\pi^0$ &$\Gamma_{total}$ &$\mathcal{B}$\\
			\hline
			&1&$\Lambda_{c1}^{ }(\frac{1}{2}^-)$          &3.70  &3.93      &7.46  &15.09    &24.52\% \\
			&2&$\Lambda_{c1}^{ }(\frac{3}{2}^-)$          &$\approx0$ &$\approx0$  &$\approx0$  &$\approx0$  &$\approx0$         \\
			&3&$\tilde\Lambda_{c0}^{ }(\frac{1}{2}^-)$    &0     &0         &0     &0       &-               \\
			&4&$\tilde\Lambda_{c1}^{ }(\frac{1}{2}^-)$    &22.22  &23.56&44.75 &90.53 &24.54\%  \\
			&5&$\tilde\Lambda_{c1}^{ }(\frac{3}{2}^-)$    &$\approx0$ &$\approx0$  &$\approx0$  &$\approx0$  &$\approx0$         \\
			&6&$\tilde\Lambda_{c2}^{ }(\frac{3}{2}^-)$    &$\approx0$ &$\approx0$  &$\approx0$  &$\approx0$  &$\approx0$        \\
			&7&$\tilde\Lambda_{c2}^{}(\frac{5}{2}^-)$     &$\approx0$ &$\approx0$  &$\approx0$  &$\approx0$  &$\approx0$         \\
			\hline
			&8   &$\tilde\Lambda_{c0}^{\rho'}(\frac{1}{2}^+)$  &$7.61\times10^{-4}$ & $9.06\times10^{-4}$  &  $6.20\times10^{-3}$  & $7.87\times10^{-3}$ &9.67\%   \\
			&9   &$\tilde\Lambda_{c0}^{\lambda'}(\frac{1}{2}^+)$  &$1.58\times10^{-3}$ &$1.88\times10^{-3}$   &  $1.30\times10^{-2}$ & $1.65\times10^{-2}$  & 9.58\%  \\
			\hline\hline
		\end{tabular}
		\label{table5}
	\end{table*}
\end{center}

\begin{center}
	\begin{table*}[t]
		\caption {Decay widths (MeV) of $\Lambda_c(2625)^+$ as $1P$-wave and $2S$-wave charmed baryons. $\mathcal{B}=\Gamma(\Sigma_c^{++}\pi^{-})/\Gamma_{total}$. }
		\begin{tabular}{lccccccc}
			\hline\hline
			&N&$\Lambda_{cJ_l} (J^P)$  &$\Sigma_c^{++}\pi^- $ &$\Sigma_c^{0}\pi^+ $ &$\Sigma_c^{+}\pi^0$ &$\Gamma_{total}$&$\mathcal{B}$  \\
			\hline
			&1&$\Lambda_{c1}^{ }(\frac{1}{2}^-)$  &19.67&19.75&21.06&60.48&32.53\% \\
			&2&$\Lambda_{c1}^{ }(\frac{3}{2}^-)$  &$0.33\times10^{-2}$&$0.33\times10^{-2}$&$0.47\times10^{-2}$ &$1.13\times10^{-2}$ &29.20\% \\
			&3&$\tilde\Lambda_{c0}^{ }(\frac{1}{2}^-)$&0  &0   &0   &0  &-     \\
			&4&$\tilde\Lambda_{c1}^{ }(\frac{1}{2}^-)$&118.01&118.50&126.35&362.86&32.52\%\\
			&5&$\tilde\Lambda_{c1}^{ }(\frac{3}{2}^-)$&$0.49\times10^{-2}$&$0.50\times10^{-2}$&$0.70\times10^{-2}$ &$1.69\times10^{-2}$&28.99\%\\
			&6&$\tilde\Lambda_{c2}^{ }(\frac{3}{2}^-)$&$0.89\times10^{-2}$&$0.90\times10^{-2}$&$0.13\times10^{-1}$ &$3.09\times10^{-2}$&28.80\% \\
			&7&$\tilde\Lambda_{c2}^{1 }(\frac{5}{2}^-)$&$0.39\times10^{-2}$&$0.40\times10^{-2}$&$0.56\times10^{-2}$ &$1.35\times10^{-2}$&28.89\%\\
			\hline
			&8   &$\tilde\Lambda_{c0}^{\rho'}(\frac{1}{2}^+)$  &$8.18\times10^{-2}$  &$8.27\times10^{-2}$ & 0.10 & 0.26 &31.46\%   \\
			&9   &$\tilde\Lambda_{c0}^{\lambda'}(\frac{1}{2}^+)$  &0.18  &0.18 & 0.23 &  0.59 &30.51\%   \\
			\hline\hline
		\end{tabular}
		\label{table6}
	\end{table*}
\end{center}

\begin{center}
	\begin{table*}[t]
		\caption{Decay widths (MeV) of $\Lambda_c(2595)^+$ as $D$-wave excitations. The branching fractions ratios $\mathcal{B}=\Gamma(\Lambda_c(2595)^+ \to \Sigma_c^{++}\pi^- )/ \Gamma_{total}$.}
		\begin{tabular}{ccc|ccccc} \hline \hline
			&N~&$\Lambda_{cJ_l}(J^P)$ &$\Sigma_c^{++}\pi^- $ &$\Sigma_c^{0}\pi^+ $ &$\Sigma_c^{+}\pi^0$ &$\Gamma_{total}$ &$\mathcal{B}$                                                \\
			\hline\hline
			&1	&$\Lambda_{c2}(\frac{3}{2}^+)$      &$4.11\times10^{-4}$  &$4.88\times10^{-4}$  &$3.36\times10^{-3}$   &$4.26\times10^{-3}$   &$9.65\%$   \\
			
			&2	&$\Lambda_{c2}(\frac{5}{2}^+)$      &$\approx0$ &$\approx0$  &$\approx0$  &$\approx0$  &$\approx0$       \\
			
			&3	&$\hat\Lambda_{c2}(\frac{3}{2}^+)$  &$3.70\times10^{-3}$  &$4.40\times10^{-3}$  &$3.02\times10^{-2}$   &$3.83\times10^{-2}$   &$9.66\%$   \\
			
			&4	&$\hat\Lambda_{c2}(\frac{5}{2}^+)$  &$\approx0$ &$\approx0$  &$\approx0$  &$\approx0$  &$\approx0$        \\
			
			&5	&$\check\Lambda_{c0}^{1}(\frac{1}{2}^+)$  &0  &0   &0  &0  &-         \\
			
			&6	&$\check\Lambda_{c1}^{1}(\frac{1}{2}^+)$  &0  &0   &0  &0  &-        \\
			
			&7	&$\check\Lambda_{c1}^{1}(\frac{3}{2}^+)$  &$\approx0$ &$\approx0$  &$\approx0$  &$\approx0$  &$\approx0$        \\
			
			&8	&$\check\Lambda_{c2}^{1}(\frac{3}{2}^+)$  &$\approx0$ &$\approx0$  &$\approx0$  &$\approx0$  &$\approx0$        \\
			
			&9	&$\check\Lambda_{c2}^{1}(\frac{5}{2}^+)$  &$\approx0$ &$\approx0$  &$\approx0$  &$\approx0$  &$\approx0$        \\
			
			&10	&$\check\Lambda_{c1}^{0}(\frac{1}{2}^+)$  &$9.08\times10^{-3}$  &$1.08\times10^{-2}$  &$7.43\times10^{-2}$ &$9.42\times10^{-2}$ &$9.64\%$  \\
			
			&11	&$\check\Lambda_{c1}^{0}(\frac{3}{2}^+)$  &$2.27\times10^{-3}$  &$2.70\times10^{-3}$  &$1.86\times10^{-2}$ &$2.36\times10^{-2}$ &$9.62\%$    \\
			
			&12	&$\check\Lambda_{c1}^{2}(\frac{1}{2}^+)$  &$1.64\times10^{-3}$  &$1.95\times10^{-3}$   &$1.34\times10^{-2}$ &$1.70\times10^{-2}$ &$9.65\%$   \\
			
			&13	&$\check\Lambda_{c1}^{2}(\frac{3}{2}^+)$  &$4.11\times10^{-4}$  &$4.88\times10^{-4}$      &$3.36\times10^{-3}$ &$4.26\times10^{-3}$ &$9.65\%$   \\
			
			&14	&$\check\Lambda_{c2}^{2}(\frac{3}{2}^+)$  &$3.70\times10^{-3}$  &$4.40\times10^{-3}$      &$3.02\times10^{-2}$ &$3.83\times10^{-2}$ &$9.66\%$   \\
			
			&15	&$\check\Lambda_{c2}^{2}(\frac{5}{2}^+)$  &$\approx0$ &$\approx0$  &$\approx0$  &$\approx0$  &$\approx0$        \\
			
			&16	&$\check\Lambda_{c3}^{2}(\frac{5}{2}^+)$  &$\approx0$ &$\approx0$  &$\approx0$  &$\approx0$  &$\approx0$        \\
			
			&17	&$\check\Lambda_{c3}^{2}(\frac{7}{2}^+)$  &$\approx0$ &$\approx0$  &$\approx0$  &$\approx0$  &$\approx0$        \\
			
			\hline\hline
		\end{tabular}
		\label{table7}
	\end{table*}
\end{center}

\begin{center}
	\begin{table*}[t]
		\caption{Decay widths (MeV) of $\Lambda_c(2625)^+$ as $D$-wave  excitations. $\mathcal{B}$ represent the ratio of branching fractions $\Gamma(\Lambda_c(2625)^+ \to \Sigma_c^{++}\pi^- )/ \Gamma_{total}$ }
		\begin{tabular}{ccc|ccccc} \hline \hline
			&N~&$\Lambda_{cJ_l}(J^P)$ &$\Sigma_c^{++}\pi^- $ &$\Sigma_c^{0}\pi^+ $ &$\Sigma_c^{+}\pi^0$ &$\Gamma_{total}$ &$\mathcal{B}$                                                 \\
			\hline\hline
			&1	&$\Lambda_{c2}(\frac{3}{2}^+)$            &$4.59\times10^{-2}$  &$4.64\times10^{-2}$  &$5.66\times10^{-2}$   &$14.89\times10^{-2}$   &$30.83\%$   \\
			&2	&$\Lambda_{c2}(\frac{5}{2}^+)$            &$\approx0$ &$\approx0$  &$\approx0$  &$\approx0$  &$\approx0$        \\
			&3	&$\hat\Lambda_{c2}(\frac{3}{2}^+)$        &0.41  &0.42  &0.51   &1.34   &$30.60\%$ \\
			&4	&$\hat\Lambda_{c2}(\frac{5}{2}^+)$        &$\approx0$ &$\approx0$  &$\approx0$  &$\approx0$  &$\approx0$         \\
			&5	&$\check\Lambda_{c0}^{1}(\frac{1}{2}^+)$  &0  &0   &0  &0  &-         \\
			&6	&$\check\Lambda_{c1}^{1}(\frac{1}{2}^+)$  &0  &0   &0  &0  &-       \\
			&7	&$\check\Lambda_{c1}^{1}(\frac{3}{2}^+)$  &$\approx0$ &$\approx0$  &$\approx0$  &$\approx0$  &$\approx0$       \\
			&8	&$\check\Lambda_{c2}^{1}(\frac{3}{2}^+)$  &$\approx0$ &$\approx0$  &$\approx0$  &$\approx0$  &$\approx0$       \\
			&9	&$\check\Lambda_{c2}^{1}(\frac{5}{2}^+)$  &$\approx0$ &$\approx0$  &$\approx0$  &$\approx0$  &$\approx0$        \\
			&10	&$\check\Lambda_{c1}^{0}(\frac{1}{2}^+)$  &1.02  &1.03  &1.25 &3.30  &$30.91\%$  \\
			&11	&$\check\Lambda_{c1}^{0}(\frac{3}{2}^+)$  &0.25  &0.26  &0.31 &0.82  &$30.49\%$  \\
			&12	&$\check\Lambda_{c1}^{2}(\frac{1}{2}^+)$  &0.18  &0.19  &0.23 &0.60  &$30.00\%$  \\
			&13	&$\check\Lambda_{c1}^{2}(\frac{3}{2}^+)$  &$4.59\times10^{-2}$  &$4.64\times10^{-2}$      &$5.66\times10^{-2}$ &$14.89\times10^{-2}$  &$30.83\%$  \\
			&14	&$\check\Lambda_{c2}^{2}(\frac{3}{2}^+)$  &0.41  &0.42  &0.51 &1.34  &$30.60\%$ \\
			&15	&$\check\Lambda_{c2}^{2}(\frac{5}{2}^+)$  &$\approx0$ &$\approx0$  &$\approx0$  &$\approx0$  &$\approx0$         \\
			&16	&$\check\Lambda_{c3}^{2}(\frac{5}{2}^+)$  &$\approx0$ &$\approx0$  &$\approx0$  &$\approx0$  &$\approx0$        \\
			&17	&$\check\Lambda_{c3}^{2}(\frac{7}{2}^+)$  &$\approx0$ &$\approx0$  &$\approx0$  &$\approx0$  &$\approx0$       \\
			\hline\hline
		\end{tabular}
		\label{table8}
	\end{table*}
\end{center}
		
From Table~\ref{table5}, $\Lambda_c(2595)^+$ seems impossibly a $~\Lambda_{c1}^{ }(\frac{3}{2}^-)$, $~\tilde\Lambda_{c0}^{ }(\frac{1}{2}^-)$, $\tilde\Lambda_{c1}^{ }(\frac{3}{2}^-)$, $\tilde\Lambda_{c2}^{ }(\frac{3}{2}^-)$, or $\tilde\Lambda_{c2}^{ }( \frac{5}{2}^-)$ for a vanishing (denoted with "0" in the table) total decay width or approximately vanishing total decay width (denoted with "$\approx 0$" in the table). It is impossibly the $\tilde\Lambda_{c1}^{ }(\frac{1}{2}^-)$ either for a large total decay width. $\Lambda_c(2595)^+$ is impossibly a $2S$-wave excitation, $\tilde\Lambda_{c0}^{}(\frac{1}{2}^+)$ or $\tilde\Lambda_{c0}^{}(\frac{1}{2}^+)$, for a much lower predicted branching fractions $B=\Gamma(\Sigma_c^{++}\pi^{-})/\Gamma_{total}$. The predicted total decay width is much smaller either in comparison with experimental data.

From Table~\ref{table7}, neither the branching ratios nor the total decay widths are consistent with experimental measurements. Therefore, $\Lambda_c(2595)^+$ is impossibly a $D$-wave excitation of $\Lambda_c$. Account for the branching fractions $B=\Gamma(\Sigma_c^{(++)}\pi^{(-)})/\Gamma_{total}$ and the total decay width, $\Lambda_c(2595)^+$ is most possibly a $1P$-wave $\Lambda_{c1}^{ }(\frac{1}{2}^-)$.

From Table~\ref{table6}, $\Lambda_c(2625)^+$ seems impossibly a $\Lambda_{c1}^{ }(\frac{1}{2}^-)$, $\tilde\Lambda_{c1}^{ }(\frac{1}{2}^-)$ or $\tilde\Lambda_{c0}^{ }(\frac{1}{2}^-)$ for a large predicted decay width or a vanishing $\Sigma_c^{++}\pi^-$ mode. For $\Lambda_c(2625)^+$, $\Sigma_c(2455)\pi$ are the only two-body decay modes of this state, and the branching fraction of the direct three-body decay mode $\Lambda_c^+\pi\pi$ is large, so it is impossible to learn this state only from the branching fraction of these two-body strong decay modes. However, the predicted masses of $\tilde\Lambda_{c1}^{ }(\frac{3}{2}^-)$, $\tilde\Lambda_{c2}^{ }(\frac{3}{2}^-)$, $\tilde\Lambda_{c2}^{}(\frac{5}{2}^-)$, the $2S$-wave excitations and the $1D$-wave excitations are much higher than that of $\Lambda_c(2625)^+$~\cite{ebert,chen3,oka}. Account for this fact, $\Lambda_c(2625)^+$ seems impossibly these charmed baryons. In short, $\Lambda_c(2625)^+$ is possibly a $P$-wave $\Lambda_{c1}^{ }(\frac{3}{2}^-)$ charmed baryon.

In the given configurations of $\Lambda_c(2595)^+$ and $\Lambda_c(2625)^+$, there is a $\lambda$-mode excitation while there is not a $\rho$-mode excitation. The two light quarks inside couple with total spin $S_\rho=0$. $\Lambda_c(2595)^+$ and $\Lambda_c(2625)^+$ make a doublet $\Lambda_{c1}(\frac{1}{2}^-,\frac{3}{2}^-)$.
		
\subsection{$\Lambda_c(2765)$ (or $\Sigma_c(2765)$)}			
			
$\Lambda_c(2765)^+$ (or $\Sigma_c(2765)^+$) is a broad state first observed in $\Lambda^+_c\pi^-\pi^+$ channel by CLEO Collaboration~\cite{prl86.4479}. However, nothing is known about its $J^P$. One even does not know whether it is a $\Lambda_c$ or a $\Sigma_c$. $\Lambda_c(2765)^+$ (or $\Sigma_c(2765)^+$) was suggested as a first orbital excitation of $\Lambda_c$ with $J^P=\frac{1}{2}^+$~\cite{capstick2}, $J^P=\frac{1}{2}^-$~\cite{zhong} or $J^P=\frac{3}{2}^+$~\cite{roberts,li}. $\Lambda_c(2765)^+$ (or $\Sigma_c(2765)^+$) was suggested as a first orbital $1P$-excitation of the $\Sigma_c$ with $J^P=\frac{1}{2}^-$~\cite{ebert} or $J^P=\frac{3}{2}^-$~\cite{ebert,garcilazo,chen3}. $\Lambda_c(2765)^+$ (or $\Sigma_c(2765)^+$) was also suggested as a first radial $2S$-excitation of $\Lambda_c$ with $J^P=\frac{1}{2}^+$ in a relativistic flux tube model~\cite{chen} and a hyper-central constituent quark model~\cite{zalak}.

In this subsection, all the possibilities of $\Lambda_{c}(2765)$ (or $\Sigma_c(2765)$) as the $1P$-wave, $2S$-wave and $1D$-wave charmed baryon with isospin $I=0$ are examined. When $\Lambda_{c}(2765)$ (or $\Sigma_c(2765)$) is assigned in these configurations, the relevant hadronic decay widths are calculated in the $^3P_0$ model and are shown in Table~\ref{table9}.

\begin{center}
\begin{table*}[htbp]
\caption{Decay widths (MeV) of $\Lambda_c(2765)^+$ as $1P$-wave and $2S$-wave charmed baryons with isospin $I=0$. $\mathcal{R}=\Gamma(\Sigma_c(2520)^{++,0}\pi^{-,+})/\Gamma(\Sigma_c(2455)^{++,0}\pi^{-,+})$.}
\begin{tabular}{lc ccccc cccc}
\hline\hline

&N&$\Lambda_{cJ_l} (J^P)~~$  &~~$\Sigma_c^{++}(2455)\pi^- $ &$~~\Sigma_c^{0}(2455)\pi^+ $ &$~~\Sigma_c^{+}(2455)\pi^0$ &$\Sigma_c^{++}(2520)\pi^- $ &$~~\Sigma_c^{0}(2520)\pi^+ $ &$~~\Sigma_c^{+}(2520)\pi^0$~~&$\Gamma_{total}$  &~~~~$\mathcal{R}$  \\
\hline
&1&$\Lambda_{c1}^{ }(\frac{1}{2}^-)$ &67.51 &67.58&68.00&0.24 &0.24  &0.27 &203.84 &0.0036  \\
&2&$\Lambda_{c1}^{ }(\frac{3}{2}^-)$ &0.62  &0.63 &0.66 &46.98&46.95 &47.74&143.58 &75.14  \\
&3&$\tilde\Lambda_{c0}^{ }(\frac{1}{2}^-)$  &0    &0    &0    &0     &0    &0   &0 &-    \\
&4&$\tilde\Lambda_{c1}^{ }(\frac{1}{2}^-)$  &405  &405  &408  &0.36  &0.36 &0.40&1219.12&0.00089   \\
&5&$\tilde\Lambda_{c1}^{ }(\frac{3}{2}^-)$  &0.94 &0.94 &0.99 &281   &281  &286 &850.87 &298.94 \\
&6&$\tilde\Lambda_{c2}^{ }(\frac{3}{2}^-)$  &1.69 &1.69 &1.79 &0.33  &0.33 &0.36&6.19   &0.20  \\
&7&$\tilde\Lambda_{c2}^{}(\frac{5}{2}^-)$   &0.75 &0.75 &0.80 &0.51  &0.51 &0.56&3.88   &0.68 \\
\hline
&8&$\tilde\Lambda_{c0}^{\rho'}(\frac{1}{2}^+)$   &1.50 &1.50  & 1.53 &  1.32& 1.32  &  1.39 &8.56&0.88                     \\
&9&$\tilde\Lambda_{c0}^{\lambda'}(\frac{1}{2}^+)$  &5.33  &5.34 & 5.53  &3.64& 3.64  & 3.88&27.36&0.68                    \\
\hline\hline
\end{tabular}
\label{table9}
\end{table*}
\end{center}

From Table~\ref{table9}, account for the fact that $\Lambda_c(2595)^+$ and $\Lambda_c(2625)^+$ have been assigned with the $\Lambda_{c1}^{ }(\frac{1}{2}^-)$ and $\Lambda_{c1}^{ }(\frac{3}{2}^-)$, respectively, $\Lambda_c(2765)^+$ (or $\Sigma_c(2765)$) seems impossibly a $P$-wave $\Lambda_{c}$. Otherwise, $\Lambda_c(2765)^+$ (or $\Sigma_c(2765)$) has an extremely small or extremely large decay width. Except for the total decay width, the strong decay behaviors of the two $2S$-wave $\tilde\Lambda_{c0}^{\rho'}(\frac{1}{2}^+)$ ($\rho$-mode excitation) and $\tilde\Lambda_{c0}^{\lambda'}(\frac{1}{2}^+)$ ($\lambda$-mode excitation) are very similar, and it is difficult to distinguish them through their strong decays. Under theoretical and experimental uncertainties, $\Lambda_c(2765)^+$ (or $\Sigma_c(2765)$) may be a $2S$-wave $\tilde\Lambda_{c0}^{\rho'}(\frac{1}{2}^+)$ or $\tilde\Lambda_{c0}^{\lambda'}(\frac{1}{2}^+)$.

When $\Lambda_{c}(2765)$ (or $\Sigma_c(2765)$) is assumed with a $1D$-wave baryon with isospin $I=0$, the relevant hadronic decay widths are calculated and presented in Table~\ref{table10}. From this table, the predicted total decay widths are around the measured one in several configurations. That is to say, $\Lambda_{c}(2765)$ (or $\Sigma_c(2765)$) is possibly a $D$-wave charmed baryons. However, one has no accurate measurement of the total decay width of $\Lambda_{c}(2765)$ (or $\Sigma_c(2765)$), and has no measurement of any branching fraction or branching ratio on its decay channel. In fact, it is not suitable to draw a confirmative conclusion in terms of such less information of $\Lambda_{c}(2765)$ (or $\Sigma_c(2765)$).
\begin{center}
	\begin{table*}[t]
		\centering
		\caption{Decay widths (MeV) of $\Lambda_c(2765)^+$ as $D$-wave excitations. $\mathcal{R}=\Gamma(\Lambda_c(2765)^+ \to \Sigma_c(2520)^{++,0}\pi^{-,+})/\Gamma(\Lambda_c(2765)^+ \to\Sigma_c(2455)^{++,0}\pi)$)}
		\begin{tabular}{ccc|cccccccc} \hline \hline
			&N &Assignment  &$\Sigma_c^{++}(2455)\pi^- $ &$\Sigma_c^{0}(2455)\pi^+ $ &$\Sigma_c^{+}(2455)\pi^0$ &$\Sigma_c^{++}(2520)\pi^- $ &$\Sigma_c^{0}(2520)\pi^+ $ &$\Sigma_c^{+}(2520)\pi^0$&$\Gamma_{total}$  &$\mathcal{R}$  \\
			\hline\hline
			& 1 &$\Lambda_{c2}(\frac{3}{2}^+)$   &1.12  &1.12  &1.15   &$8.51\times10^{-2}$  &$8.50\times10^{-2}$&$9.01\times10^{-2}$ &3.65 &0.074 \\
			& 2&$\Lambda_{c2}(\frac{5}{2}^+)$   &$4.20\times10^{-3}$  &$4.23\times10^{-3}$   &$4.56\times10^{-3}$   &0.51&0.51 &0.54&1.57 &121.00\\
			& 3 &$\hat\Lambda_{c2}(\frac{3}{2}^+)$  &10.09 &10.12 &10.43  &0.77  &0.76&0.81&32.98  &0.076 \\
			& 4 &$\hat\Lambda_{c2}(\frac{5}{2}^+)$   &$3.78\times10^{-2}$ &$3.81\times10^{-2}$   &$4.17\times10^{-2}$   &4.55   &4.55&4.82&14.04 &119.89 \\
			& 5 &$\check\Lambda_{c0}^{1}(\frac{1}{2}^+)$  &0      &0      &0      &0      &0      &0     &0    &-       \\
			& 6 &$\check\Lambda_{c1}^{1}(\frac{1}{2}^+)$  &0      &0      &0      &0      &0      &0     &0    &-       \\
			& 7 &$\check\Lambda_{c1}^{1}(\frac{3}{2}^+)$  &$\approx0$ &$\approx0$  &$\approx0$  &$\approx0$  &$\approx0$   &$\approx0$ &$\approx0$ &$\approx0$       \\
			& 8 &$\check\Lambda_{c2}^{1}(\frac{3}{2}^+)$  &$\approx0$ &$\approx0$  &$\approx0$  &$\approx0$  &$\approx0$   &$\approx0$ &$\approx0$ &$\approx0$       \\
			& 9 &$\check\Lambda_{c2}^{1}(\frac{5}{2}^+)$   &$\approx0$ &$\approx0$  &$\approx0$  &$\approx0$  &$\approx0$   &$\approx0$ &$\approx0$ &$\approx0$       \\
			& 10&$\check\Lambda_{c1}^{0}(\frac{1}{2}^+)$ &24.96 &25.03  &25.78 &4.67  & 4.67 &4.95    &90.06 &0.19\\
			& 11&$\check\Lambda_{c1}^{0}(\frac{3}{2}^+)$ &6.24 &6.26  &6.44 &11.69  &11.67  &12.37    &54.67 &1.87 \\
			& 12&$\check\Lambda_{c1}^{2}(\frac{1}{2}^+)$ &4.49 &4.50  &4.63 &0.84  &0.84  &0.89    &16.19 &0.19 \\
			& 13&$\check\Lambda_{c1}^{2}(\frac{3}{2}^+)$ &1.12  &1.12  &1.16   &2.11  &2.10&2.23 &9.84 &1.88 \\
			& 14&$\check\Lambda_{c2}^{2}(\frac{3}{2}^+)$ &10.09 &10.12 &10.43  &0.77  &0.76&0.81&32.98 &0.076 \\
			& 15&$\check\Lambda_{c2}^{2}(\frac{5}{2}^+)$ &$1.68\times10^{-2}$   &$1.69\times10^{-2}$   &$1.83\times10^{-2}$   &4.55  &4.55 &4.82 &13.97 &270.03\\
			& 16&$\check\Lambda_{c3}^{2}(\frac{5}{2}^+)$  &$1.92\times10^{-2}$  &$1.93\times10^{-2}$   &$2.09\times10^{-2}$   &$2.60\times10^{-3}$   &$2.59\times10^{-3}$   &$2.98\times10^{-3}$  &$6.72\times10^{-2}$   &0.13\\
			& 17&$\check\Lambda_{c3}^{2}(\frac{7}{2}^+)$  &$1.08\times10^{-2}$  &$1.09\times10^{-2}$   &$1.17\times10^{-2}$        &$3.51\times10^{-3}$  &$3.50\times10^{-3}$ &$4.03\times10^{-3}$ &$4.44\times10^{-2}$   &0.32 \\
			\hline\hline
		\end{tabular}
		\label{table10}
	\end{table*}
\end{center}

\subsection{$\Lambda_c(2860)$, $\Lambda_c(2880)$ and $\Lambda_c(2940)$}	

$\Lambda_c(2860)^+$ as a newly reported $\Lambda_c$ baryon was first observed by the LHCb Collaboration in the $D^0p$ channel~\cite{R.Aaij}. The mass and width of $\Lambda_c(2860)^+$ were measured. The mass of $\Lambda_c(2860)^+$ is consistent with the predictions for an orbital $D$-wave $\Lambda_c$
excitation with $J^P={3\over 2}^+$~\cite{zhu3,chen3}. In particular, quantum numbers of $\Lambda_c(2860)^+$ were found to be $J^P={\frac{3}{2}}^+$, the other quantum numbers were excluded with a significance of more than $6$ standard deviations~\cite{R.Aaij}.

$\Lambda_c(2880)^+$ was first observed by the CLEO Collaboration in $\Lambda_c^+\pi^-\pi^+$~\cite{prl86.4479} and confirmed by the BaBar Collaboration in the $D^0p$ channel~\cite{babar2007}. From an analysis of angular distributions in $\Lambda_c(2880)^+\to \Sigma_c(2455)^{0,++}\pi^{+,-}$ decays and the measured $R=\Gamma(\Sigma_c(2520)\pi)/\Gamma(\Sigma_c(2455)\pi)=0.225\pm0.062\pm 0.0255$, the preferred quantum numbers of $\Lambda_c(2880)^+$ state were constrained to $J^P={\frac{5}{2}}^+$ by Belle Collaboration~\cite{prl98.262001}. Recently, the LHCb Collaboration studied the spectrum of excited $\Lambda_{c}$ states that decay into $D^0p$ channel and measured the mass and width of $\Lambda_c(2880)^+$. The preferred spin of $\Lambda_c(2880)^+$ is found to be ${5\over 2}$, and the spin assignments ${1\over 2}$ and ${3\over 2}$ were excluded~\cite{R.Aaij}.

$\Lambda_c(2940)^+$ was first observed by the BaBar Collaboration in $D^0p$ invariant mass distribution~\cite{babar2007}. The spin-parity of $\Lambda_c(2940)^+$ was constrained to $J^P={\frac{3}{2}}^-$ by LHCb Collaboration~\cite{R.Aaij} though other solutions with spins ${1\over 2}$ to ${7\over 2}$ cannot be excluded.

$\Lambda_c(2860)^+$ was assigned with a $D$-wave charmed baryon with $J^P={3\over 2}^+$~\cite{chen2,npb926.(2018)467}. In particular, $\Lambda_c(2860)^+$ and $\Lambda_c(2880)^+$ are supposed to form a $D$-wave doublet $[{3\over 2}^+,{5\over 2}^+]$~\cite{chen2}.

$\Lambda_c(2880)^+$ was once assigned with quantum numbers $J^P={1\over 2}^-$ or $J^P={3\over 2}^-$~\cite{garcilazo}, it was also assigned as a $D$-wave state with $J^P={3\over 2}^+$~\cite{zhong,zhong2}. In most references~\cite{cheng2,vijande,zhu,zhu2,zhu3,zhu4,ebert,chen,chen2,chen3,oka}, $\Lambda_c(2880)^+$ was conjectured as an excited charmed baryon with $J^P={5\over 2}^+$ though its structure may be different in these references.

In addition to an $S$-wave $D^*N$ molecular state interpretation~\cite{he,dong1,dong2,dong3,liu,ortega,jianrong,ping}, $\Lambda_c(2940)^+$ was interpreted as an excited charmed baryon with different $J^P$ quantum numbers as shown in Table.~\ref{table1}.

In order to check all the possibilities as charmed baryons candidates, $\Lambda_c(2860)^+$, $\Lambda_c(2880)$ and $\Lambda_c(2940)$ are studied as the $1P$-wave, $2S$-wave and $1D$-wave states in detail in the $^3P_0$ model. Their OZI-allowed two-body strong decay channels are all given and relevant decay widths have been estimated. Their decay widths as $1P$-wave and $2S$-wave charmed baryons are presented in Tables.~\ref{table11}, ~\ref{table12} and ~\ref{table13}. Their decay widths as $1D$-wave charmed baryons are presented in Tables.~\ref{table14}, ~\ref{table15} and ~\ref{table16}.

\begin{center}
\begin{table*}[t]
\caption{Decay widths (MeV) of $\Lambda_c(2860)^+$ as $1P$-wave and $2S$-wave charmed baryons. $\mathcal{R}=\Gamma(\Sigma_c(2520)^{++,0}\pi)/\Gamma(\Sigma_c(2455)^{++,0}\pi)$.}
\begin{tabular}{lcc|cccccccccc}
\hline\hline
&N &Assignment  &$D^0P$  &$D^+N$ &$\Sigma_c^{++,0}\pi^{-,+}$  &$\Sigma_c^{+}\pi^0$ &$\Sigma_c^{*++,0}\pi^{-,+}$&$\Sigma_c^{*+}\pi^0$ &$\Gamma_{total}$ &$\mathcal{B1}$ &$\mathcal{B2}$ &$\mathcal{R}$ \\
\hline
&1&$\Lambda_{c1}^{ }(\frac{1}{2}^-)$    &0   &0    &176.75    &88.46  &4.15  &2.18  &271.54 &65.09\% &1.52\%  &0.02          \\
&2&$\Lambda_{c1}^{ }(\frac{3}{2}^-)$    &0   &0    &5.93     &3.07   &151.66 &76.23 &236.89 &2.50\%  &64.02\% &25.57        \\
&3&$\tilde\Lambda_{c0}^{ }(\frac{1}{2}^-)$ &726.04    &703.59   &0     &0         &0    &0      &1429.63    &-  &-  &-  \\
&4&$\tilde\Lambda_{c1}^{ }(\frac{1}{2}^-)$ &0   &0    &1060.53&530.77  &6.22     &3.27 &1600.79 &66.25\%  &0.38\% &0.0058         \\
&5&$\tilde\Lambda_{c1}^{ }(\frac{3}{2}^-)$ &0   &0    &8.90  &4.60    &900.66   &452.50&1366.66 &0.65\%   &65.90\%  &101.20        \\
&6&$\tilde\Lambda_{c2}^{ }(\frac{3}{2}^-)$ &2.48&1.84 &16.02 &8.29    &5.60  &2.94 &37.17     &43.10\% &15.07\%   &0.35       \\
&7&$\tilde\Lambda_{c2}^{}(\frac{5}{2}^-)$  &2.48&1.84 &7.13  &3.68    &8.72  &4.58 &28.43 &25.08\%    &30.67\%  &1.22           \\
\hline
&8&$\tilde\Lambda_{c0}^{\rho'}(\frac{1}{2}^+)$   &0  &0   &5.17    &2.60  &7.45  &3.78 &19.00  &27.21\% &39.21\% &1.44          \\
&9&$\tilde\Lambda_{c0}^{\lambda'}(\frac{1}{2}^+)$  &0  &0   &30.09 &15.37  &29.89  &15.39 &90.74  &40.66\% &32.95\% &0.99         \\
\hline\hline
\end{tabular}
\label{table11}
\end{table*}
\end{center}

\begin{center}
\begin{table*}[h]
\caption{Decay widths (MeV) of $\Lambda_c(2880)^+$ as $1P$-wave and $2S$-wave charmed baryons. The branching fractions $\mathcal{B}1=\Gamma(\Lambda_c(2880)^+ \to \Sigma_c(2455)^{++,0}\pi^{-,+} )/ \Gamma_{total}$ and $\mathcal{B}2=\Gamma(\Lambda_c(2880)^+ \to \Sigma_c(2520)^{++,0}\pi^{-,+} )/ \Gamma_{total}$, respectively. $\mathcal{R}$ represents the ratio of $\mathcal{B}2/\mathcal{B}1$.}
\begin{tabular}{lcc|cccccccccc}
\hline\hline
&N &Assignment  &$D^0P$  &$D^+N$ &$\Sigma_c^{++,0}\pi^{-,+}$  &$\Sigma_c^{+}\pi^0$ &$\Sigma_c^{*++,0}\pi^{-,+}$&$\Sigma_c^{*+}\pi^0$ &$\Gamma_{total}$ &$\mathcal{B}1$ &$\mathcal{B}2$ &$\mathcal{R}$ \\
\hline
&1&$\Lambda_{c1}^{ }(\frac{1}{2}^-)$       &0    &0    &184.10 &92.04 &6.53 &3.41     &286.08  &64.35\%  &2.28\%  &0.04     \\
&2&$\Lambda_{c1}^{ }(\frac{3}{2}^-)$       &0    &0    &8.45  &4.35  &165.42  &83.02  &261.24  &3.23\%   &63.32\% &19.60      \\
&3&$\tilde\Lambda_{c0}^{ }(\frac{1}{2}^-)$ &784.33  &776.99  &0   &0   &0  &0     &1561.32&-  &-& -   \\
&4&$\tilde\Lambda_{c1}^{ }(\frac{1}{2}^-)$ &0     &0   &1104.59 &552.21 &9.80 &5.11  &1671.71   &66.07\%  &0.59\%  &0.0088     \\
&5&$\tilde\Lambda_{c1}^{ }(\frac{3}{2}^-)$ &0     &0   &12.68  &6.53  &977.79  &490.44 &1487.44 &0.85\% &65.74\%   &77.34      \\
&6&$\tilde\Lambda_{c2}^{ }(\frac{3}{2}^-)$ &6.53  &5.38&22.82  &11.76  &8.82 &4.60   &59.91 &38.09\% &14.72\% &0.39   \\
&7&$\tilde\Lambda_{c2}^{}(\frac{5}{2}^-)$  &6.53 &5.38 &10.18  &5.23  &13.72 &7.15   &48.19 &21.12\% &28.47\% &1.35    \\
\hline
&8&$\tilde\Lambda_{c0}^{\rho'}(\frac{1}{2}^+)$&  0  &0  &5.48  &2.74   &8.78  &4.44 &21.44&25.56\%   &40.95\% &1.60         \\
&9&$\tilde\Lambda_{c0}^{\lambda'}(\frac{1}{2}^+)$  &0  &0  &38.16  &19.45  &40.39   &20.73 &118.73 &32.14\%   &34.02\% &1.06      \\
\hline\hline
\end{tabular}
\label{table12}
\end{table*}
\end{center}

\begin{center}
\begin{table*}[h]
\caption{Decay widths (MeV) of $\Lambda_c(2940)^+$ as $1P$-wave and $2S$-wave charmed baryons. $\mathcal{R}=B(\Lambda_{c}(2940)^{+} \to \Sigma_c{(2520)} \pi^{}/B(\Lambda_{c}(2940)^{+} \to \Sigma_c{(2455)} \pi^{}$).}
\begin{tabular}{lcc|ccccccccccc}
\hline\hline
&N &Assignment  &$D^0P$  &$D^+N$ &$\Sigma_c^{++,0}\pi^{-,+}$  &$\Sigma_c^{+}\pi^0$ &$\Sigma_c^{*++,0}\pi^{-,+}$&$\Sigma_c^{*+}\pi^0$&$\Sigma_c^{'++,0}\pi^{-,+}$ &$\Sigma_c^{'+}\pi^0$ &$\Sigma_c^{''++}\pi^{0}$  &$\Gamma_{total}$ &$\mathcal{R}$ \\
\hline
&1&$\Lambda_{c1}^{ }(\frac{1}{2}^-)$     &0  &0  &192.45  &96.01 &15.67  &8.08  &0   &0 &0  &312.21 &0.08\\
&2&$\Lambda_{c1}^{ }(\frac{3}{2}^-)$     &0  &0  &17.06 &8.73  &191.06 &95.66 &0   &0 &0  &312.51 &11.20    \\
&3&$\tilde\Lambda_{c0}^{ }(\frac{1}{2}^-)$   &768.52  &777.42 &0  &0 &0 &0 &0.38 &0.22 &0.07&1546.61  & -   \\
&4&$\tilde\Lambda_{c1}^{ }(\frac{1}{2}^-)$   &0  &0  &1154.72&576.08 &23.51 &12.12 &2.17 &1.29 &0.10  &1769.99  &0.02\\
&5&$\tilde\Lambda_{c1}^{ }(\frac{3}{2}^-)$   &0  &0  &25.59  &13.09  &1111.09 &555.78 &0.54   &0.32 &0.24&1706.65 &43.42  \\
&6&$\tilde\Lambda_{c2}^{ }(\frac{3}{2}^-)$   &25.15  &22.64 &46.04  &23.57 &21.15 &10.91  &1.38 &0.82  &0.21  &151.87 &0.46\\
&7&$\tilde\Lambda_{c2}^{}(\frac{5}{2}^-)$    &25.15  &22.64 &20.47   &10.47  &32.90 &16.97 &1.16 &0.68  &0.23 &130.67 &1.61\\
\hline
&8&$\tilde\Lambda_{c0}^{\rho'}(\frac{1}{2}^+)$  &0   &0   &5.32 &2.63  &10.88   &5.44 &0  &0 &0 &24.27  &2.05    \\
&9&$\tilde\Lambda_{c0}^{\lambda'}(\frac{1}{2}^+)$  &0  &0   &61.37 &31.15  &72.62   &37.06 &0  &0 &0 &202.20  &1.18     \\
\hline\hline
\end{tabular}
\label{table13}
\end{table*}
\end{center}

\begin{center}
	\begin{table*}[h]
		\centering
		\caption{Decay widths (MeV) of $\Lambda_c(2860)^+$ as $D$-wave excitations. The branching ratios $\mathcal{B}1=\Gamma(\Lambda_c(2860)^+ \to \Sigma_c(2455)^{++,0}\pi^{-,+} )/ \Gamma_{total}$ and $\mathcal{B}2=\Gamma(\Lambda_c(2860)^+ \to \Sigma_c(2520)^{++,0}\pi^{-,+} )/ \Gamma_{total}$, respectively. $\mathcal{R}$ represents the ratio of $\mathcal{B}2/\mathcal{B}1$.}
		\begin{tabular}{ccc|cccccccccc} \hline \hline
			&N &Assignment  &$D^0P$  &$D^+N$ &$\Sigma_c^{++,0}\pi^{-,+}$  &$\Sigma_c^{+}\pi^0$ &$\Sigma_c^{*++,0}\pi^{-,+}$&$\Sigma_c^{*+}\pi^0$ &$\Gamma_{total}$ &$\mathcal{B}1$ &$\mathcal{B}2$ &$\mathcal{R}$ \\
			\hline\hline
			&1 &$\Lambda_{c2}(\frac{3}{2}^+)$ &0 &0 &5.38 &2.74  &0.64 &0.33&9.09&59.19\%  &7.04\%  &0.12\\  %01
			&2 &$\Lambda_{c2}(\frac{5}{2}^+)$ &0 &0 &$6.92\times10^{-2}$&$3.63\times10^{-2}$  &3.63 &1.86  &5.60 &1.24\% &64.82\% &52.27 \\  %02
			&3 &$\hat\Lambda_{c2}(\frac{3}{2}^+)$ &0 &0   &48.46&24.62 &5.70    &2.92   &81.70 &59.31\%  &6.98\% &0.12 \\  %03
			&4 &$\hat\Lambda_{c2}(\frac{5}{2}^+)$ &0 &0  &0.62  &0.33  &32.67  &16.76    &50.38 &1.23\% &64.85\% &52.72 \\  %04
			&5 &$\check\Lambda_{c0}^{1}(\frac{1}{2}^+)$&0 &0  &0  &0  &0 &0  &0   &-  &- &-     \\  %05
			&6 &$\check\Lambda_{c1}^{1}(\frac{1}{2}^+)$  &0 &0 &0  &0  &$\approx0$  &$\approx0$ &$\approx0$ &-   &-  &-    \\
			&7 &$\check\Lambda_{c1}^{1}(\frac{3}{2}^+)$ &0&0  &$\approx0$  &$\approx0$  &0  &0  &$\approx0$   &-   &- &-  \\  %07
			&8 &$\check\Lambda_{c2}^{1}(\frac{3}{2}^+)$ &$\approx0$ &$\approx0$  &$\approx0$  &$\approx0$  &$\approx0$ &0  &$\approx0$  &-  &- &-  \\  %08
			&9 &$\check\Lambda_{c2}^{1}(\frac{5}{2}^+)$ &$\approx0$ &$\approx0$ &$\approx0$ &$\approx0$  &$\approx0$  &$\approx0$  &$\approx0$   &-   &-  &-   \\  %09
			&10&$\check\Lambda_{c1}^{0}(\frac{1}{2}^+)$ &21.52  &18.35 &120.40  &61.19 &33.57  &17.22  &272.25 &44.22\% &12.33\% &0.28\\  %10
			&11&$\check\Lambda_{c1}^{0}(\frac{3}{2}^+)$ &21.52  &18.35 &30.10 &15.30 &82.92  &43.05 &211.24 &14.25\% &39.25\% &2.75 \\  %11
			&12&$\check\Lambda_{c1}^{2}(\frac{1}{2}^+)$ &15.48  &13.21 &21.54  &10.94 &6.02  &3.09  &70.28 &30.65\% &8.57\%  &0.28 \\ %12
			&13&$\check\Lambda_{c1}^{2}(\frac{3}{2}^+)$ &15.48  &13.21 &5.38  &2.74 &15.07  &7.73  &59.61  &9.03\%  &25.28\% &2.80 \\  %13
			&14&$\check\Lambda_{c2}^{2}(\frac{3}{2}^+)$ &0     &0    &48.45  &24.62 &5.54  &2.85  &81.46  &59.48\% &6.80\%  &0.11\\  %14
			&15&$\check\Lambda_{c2}^{2}(\frac{5}{2}^+)$ &0     &0    &0.28  &0.15   &32.60  &16.72  &49.75&0.56\% &65.53\% &117.02\\  %15
			&16&$\check\Lambda_{c3}^{2}(\frac{5}{2}^+)$  &$3.45\times10^{-2}$  &$2.27\times10^{-2}$  &0.32&0.17   &$9.55\times10^{-2}$ &$5.10\times10^{-2}$    &0.69  &46.38\% &13.84\% &0.30\\  %16
			&17&$\check\Lambda_{c3}^{2}(\frac{7}{2}^+)$  &$3.45\times10^{-2}$  &$2.27\times10^{-2}$  &0.18   &0.09 &0.13    &0.07  &0.53  &33.96\% &24.53\% &0.72\\  %17
			\hline\hline
		\end{tabular}
		\label{table14}
	\end{table*}
\end{center}

\begin{center}
	\begin{table*}[h]
		\centering
		\caption{Decay widths (MeV) of $\Lambda_c(2880)^+$ as $D$-wave excitations. The branching ratios $\mathcal{B}1=\Gamma(\Lambda_c(2880)^+ \to \Sigma_c(2455)^{++,0}\pi^{-,+} )/ \Gamma_{total}$ and $\mathcal{B}2=\Gamma(\Lambda_c(2880)^+ \to \Sigma_c(2520)^{++,0}\pi^{-,+} )/ \Gamma_{total}$, respectively. $\mathcal{R}$ represents the ratio of $\mathcal{B}2/\mathcal{B}1$.}
		\begin{tabular}{ccc|cccccccccc} \hline \hline
			&N &Assignment  &$D^0P$  &$D^+N$ &$\Sigma_c^{++,0}\pi^{-,+}$  &$\Sigma_c^{+}\pi^0$ &$\Sigma_c^{*++,0}\pi^{-,+}$&$\Sigma_c^{*+}\pi^0$ &$\Gamma_{total}$ &$\mathcal{B}1$ &$\mathcal{B}2$ &$\mathcal{R}$ \\
			\hline\hline
			&1 &$\Lambda_{c2}(\frac{3}{2}^+)$ &0 &0 &6.48 &3.28  &0.84 &0.43&11.03&58.75\%  &7.62\%  &0.13\\  %01
			&2 &$\Lambda_{c2}(\frac{5}{2}^+)$ &0 &0 &$11.16\times10^{-2}$&$5.82\times10^{-2}$  &4.70  &2.40  &7.27 &1.54\% &64.65\% &41.98 \\  %02
			&3 &$\hat\Lambda_{c2}(\frac{3}{2}^+)$ &0 &0   &58.30 &29.55 &7.50    &3.84   &99.19 &58.78\% &7.56\% &0.13 \\  %03
			&4 &$\hat\Lambda_{c2}(\frac{5}{2}^+)$ &0 &0  &1.00  &0.52  &42.27  &21.59   &65.38  &1.53\% &64.65\% &42.25 \\  %04
			&5 &$\check\Lambda_{c0}^{1}(\frac{1}{2}^+)$&0 &0  &0  &0  &0 &0  &0   &-  &- &-     \\  %05
			&6 &$\check\Lambda_{c1}^{1}(\frac{1}{2}^+)$  &0 &0 &$\approx0$  &0  &$\approx0$  &$\approx0$ &$\approx0$ &-   &-  &-    \\
			&7 &$\check\Lambda_{c1}^{1}(\frac{3}{2}^+)$ &0&0  &$\approx0$  &$\approx0$  &0  &0  &$\approx0$   &-   &- &-  \\  %07
			&8 &$\check\Lambda_{c2}^{1}(\frac{3}{2}^+)$ &$\approx0$ &$\approx0$  &$\approx0$  &$\approx0$  &$\approx0$ &$\approx0$  &$\approx0$  &-  &- &-  \\  %08
			&9 &$\check\Lambda_{c2}^{1}(\frac{5}{2}^+)$ &0 &0 &$\approx0$ &$\approx0$  &$\approx0$  &0  &$\approx0$   &-   &-  &-   \\  %09
			&10&$\check\Lambda_{c1}^{0}(\frac{1}{2}^+)$ &35.56  &32.29&145.13  &73.57 &43.43 &22.18 &352.16 &41.21\% &12.33\% &0.30\\  %10
			&11&$\check\Lambda_{c1}^{0}(\frac{3}{2}^+)$ &35.56  &32.29 &36.28  &18.39 &108.57  &55.45  &286.54 &12.66\% &37.89\% &2.99\\  %11
			&12&$\check\Lambda_{c1}^{2}(\frac{1}{2}^+)$ &25.50  &23.17&25.92  &13.13 &7.78  &3.98  &99.48  &26.06\% &7.82\%  &0.30 \\ %12
			&13&$\check\Lambda_{c1}^{2}(\frac{3}{2}^+)$ &25.50  &23.17&6.48  &3.28 &19.46  &9.94  &87.83  &7.38\%  &22.16\% &3.00 \\  %13
			&14&$\check\Lambda_{c2}^{2}(\frac{3}{2}^+)$ &0     &0    &58.30  &29.55 &7.23  &3.70  &98.78  &59.02\% &7.32\%  &0.12\\  %14
			&15&$\check\Lambda_{c2}^{2}(\frac{5}{2}^+)$ &0     &0    &0.44  &0.23   &42.15  &21.52  &64.34 &0.68\% &65.51\% &96.34\\  %15
			&16&$\check\Lambda_{c3}^{2}(\frac{5}{2}^+)$  &0.13  &0.10  &0.51&0.27  &0.18 &0.09   &1.28  &39.84\% &14.06\% &0.35\\  %16
			&17&$\check\Lambda_{c3}^{2}(\frac{7}{2}^+)$  &0.13  &0.10  &0.28   &0.15 &0.24    &0.13&1.03  &27.18\% &23.30\% &0.86\\  %17
			\hline\hline
		\end{tabular}
		\label{table15}
	\end{table*}
\end{center}

\begin{center}
	\begin{table*}[h]
		\centering
		\caption{Decay widths (MeV) of $\Lambda_c(2940)^+$ as $D$-wave excitations. $\mathcal{R}=\Gamma(\Sigma_c(2520)^{++.0}\pi)/\Gamma(\Sigma_c(2455)^{++.0}\pi)$}
		\begin{tabular}{ccc|ccccccccccc} \hline \hline
			&N &Assignment  &$D^0P$  &$D^+N$ &$\Sigma_c^{++,0}\pi^{-,+}$  &$\Sigma_c^{+}\pi^0$ &$\Sigma_c^{*++,0}\pi^{-,+}$&$\Sigma_c^{*+}\pi^0$&$\Sigma_c^{'++,0}\pi^{-,+}$ &$\Sigma_c^{'+}\pi^0$ &$\Sigma_c^{''++}\pi^{0}$  &$\Gamma_{total}$ &$\mathcal{R}$ \\
			\hline\hline
			&1 &$\Lambda_{c2}(\frac{3}{2}^+)$ &0 &0 &9.14 &4.61  &1.42 &0.73 &0 &0 &0&15.90 &0.16\\  %01
			&2 &$\Lambda_{c2}(\frac{5}{2}^+)$ &0 &0 &0.28   &0.15  &7.56    &3.84   &0&0 &0 &11.83&27.00 \\  %02
			&3 &$\hat\Lambda_{c2}(\frac{3}{2}^+)$ &0 &0  &82.25 &41.51 &12.86 &6.55 &0 &0 &0&143.17&0.16\\%03
			&4 &$\hat\Lambda_{c2}(\frac{5}{2}^+)$ &0 &0  &2.60 &1.34 &68.12  &34.56   &0 &0 &0 &106.62&26.20\\  %04
			&5 &$\check\Lambda_{c0}^{1}(\frac{1}{2}^+)$&0 &0  &0  &0  &0 &0  &0   &0  &0&0 &-     \\  %05
			&6 &$\check\Lambda_{c1}^{1}(\frac{1}{2}^+)$  &0 &0   &0 &0  &$\approx0$  &$\approx0$ &47.46 &24.94 &$1.83\times10^{-4}$  &72.40 &-   \\  %06
			&7 &$\check\Lambda_{c1}^{1}(\frac{3}{2}^+)$ &0 &0  &$\approx0$  &$\approx0$  &0 &0  &$3.00\times10^{-3}$ &$1.99\times10^{-3}$   &12.46 &12.47 &-  \\  %07
			&8 &$\check\Lambda_{c2}^{1}(\frac{3}{2}^+)$ &0 &0  &$\approx0$  &$\approx0$  &$\approx0$  &$\approx0$ &$5.40\times10^{-3}$   &$3.58\times10^{-3}$  &$1.65\times10^{-4}$  &$9.15\times10^{-3}$ &- \\  %08
			&9 &$\check\Lambda_{c2}^{1}(\frac{5}{2}^+)$ &$\approx0$ &$\approx0$  &$\approx0$  &$\approx0$  &$\approx0$ &$\approx0$ &$2.40\times10^{-3}$  &$1.59\times10^{-3}$   &$2.57\times10^{-4}$   &$4.25\times10^{-3}$  &-  \\  %09
			&10&$\check\Lambda_{c1}^{0}(\frac{1}{2}^+)$ &65.96  &63.23  &205.65 &103.81 &69.87  &35.44  &$1.60\times10^{-1}$ &$8.75\times10^{-2}$ &$7.61\times10^{-4}$ &544.21&0.34\\  %10
			&11&$\check\Lambda_{c1}^{0}(\frac{3}{2}^+)$ &65.96  &63.23  &51.41 &25.95  &174.67  &88.59 &$1.24\times10^{-2}$&$8.17\times10^{-3}$ &$3.37\times10^{-2}$ &469.86&3.40\\  %11
			&12&$\check\Lambda_{c1}^{2}(\frac{1}{2}^+)$ &46.93  &45.02  &36.56  &18.45 &12.48 &6.33&24.64 &12.90&$2.30\times10^{-4}$&203.31 &0.34\\ %12
			&13&$\check\Lambda_{c1}^{2}(\frac{3}{2}^+)$ &46.93  &45.02   &9.14 &4.61 &31.20 &15.82 &$3.72\times10^{-3}$&$2.46\times10^{-3}$&6.60&159.33&3.41 \\  %13
			&14&$\check\Lambda_{c2}^{2}(\frac{3}{2}^+)$ &0   &0   &82.25  &41.51 &11.96  &6.07  &$2.14\times10^{-3}$&$1.42\times10^{-3}$ &$6.54\times10^{-5}$ &141.79&0.15\\  %14
			&15&$\check\Lambda_{c2}^{2}(\frac{5}{2}^+)$ &0 &0 &1.16 &0.60  &67.72  &34.35  &$9.53\times10^{-4}$&$6.32\times10^{-4}$ &$1.02\times10^{-4}$&103.83&58.38\\  %15
			&16&$\check\Lambda_{c3}^{2}(\frac{5}{2}^+)$  &0.88 &0.76  &1.32 &0.68   &0.58 &0.30   &$10.04\times10^{-4}$ &$6.62\times10^{-4}$ &$8.96\times10^{-6}$ &4.52&0.44\\  %16
			&17&$\check\Lambda_{c3}^{2}(\frac{7}{2}^+)$  &0.88 &0.76  &0.74 &0.38   &0.78 &0.40   &$3.28\times10^{-7}$ &$2.72\times10^{-7}$ &$4.05\times10^{-5}$ &3.94&1.05\\  %17
			\hline\hline
		\end{tabular}
		\label{table16}
	\end{table*}
\end{center}

From Table~\ref{table11} and Table~\ref{table14}, there are two $P$-wave assignments ($\tilde\Lambda_{c2}^{ }(\frac{3}{2}^-)$ and $\tilde\Lambda_{c2}^{ }(\frac{5}{2}^-)$) suitable for $\Lambda_c(2860)^+$, which has an observable $D^0p$ mode and a comparable $\Gamma_{total}$ with experiment. There are also two $D$-wave assignments ($\check\Lambda_{c1}^{2}(\frac{1}{2}^+)$ or $\check\Lambda_{c1}^{2}(\frac{3}{2}^+)$) suitable for $\Lambda_c(2860)^+$ for the same reason. If the experimental constraint $J^P={3\over 2}^+$ for $\Lambda_c(2860)^+$ is true~\cite{R.Aaij}, then $\Lambda_c(2860)^+$ is only possibly the $D$-wave $\check\Lambda_{c1}^{2}(\frac{3}{2}^+)$. In this case, the branching ratio $R=\Gamma(\Sigma_c(2520)^{++,0}\pi^{-,+})/\Gamma(\Sigma_c(2455)^{++,0}\pi^{-,+})=2.8$, and $\Lambda_c(2860)^+$ has a total decay $\Gamma=59.6$ MeV. For the purpose of identification of $\Lambda_c(2860)^+$, it is very important to measure the branching ratio $R=\Gamma(\Sigma_c(2520)^{++,0}\pi^{-,+})/\Gamma(\Sigma_c(2455)^{++,0}\pi^{-,+})$.

From Table~\ref{table12}, the observation of a $D^0p$ mode, the measured branching ratio $R=\Gamma(\Sigma_c(2520)\pi)/\Gamma(\Sigma_c(2455)\pi)=0.225\pm0.062\pm 0.0255$ and the total decay width indicate that $\Lambda_c(2880)^+$ is impossibly a $1P$-wave or $2S$-wave charmed baryon (the $1P$-wave $\tilde\Lambda_{c2}^{ }(\frac{3}{2}^-)$ assignment has a comparable $R$ but a much larger predicted total decay width in comparison with experimental data). From Table~\ref{table15}, the observation of a $D^0p$ mode and the measured $R=\Gamma(\Sigma_c(2520)\pi)/\Gamma(\Sigma_c(2455)\pi)=0.225\pm0.062\pm 0.0255$ indicate that $\Lambda_c(2880)^+$ may be a $\check\Lambda_{c1}^{0}(\frac{1}{2}^+)$, $\check\Lambda_{c1}^{2}(\frac{1}{2}^+)$ or $\check\Lambda_{c3}^{2}(\frac{5}{2}^+)$. Account for the much larger predicted total decay widths of  $\check\Lambda_{c1}^{0}(\frac{1}{2}^+)$ and $\check\Lambda_{c1}^{2}(\frac{1}{2}^+)$ in comparison with experiment, $\Lambda_c(2880)^+$ is possibly the $D$-wave $\check\Lambda_{c3}^{2}(\frac{5}{2}^+)$.

From Table~\ref{table13} and Table~\ref{table16}, there are three $P$-wave assignments ($\tilde\Lambda_{c0}^{ }(\frac{1}{2}^-)$, $\tilde\Lambda_{c2}^{ }(\frac{3}{2}^-)$, $\tilde\Lambda_{c2}^{ }(\frac{5}{2}^-)$) and six $D$-wave assignments ($\check\Lambda_{c1}^{0}(\frac{1}{2}^+)$, $\check\Lambda_{c1}^{0}(\frac{3}{2}^+)$, $\check\Lambda_{c1}^{2}(\frac{1}{2}^+)$, $\check\Lambda_{c1}^{2}(\frac{3}{2}^+)$, $\check\Lambda_{c3}^{2}(\frac{5}{2}^+)$ and $\check\Lambda_{c3}^{2}(\frac{7}{2}^+)$), which have an observable $D^0p$ mode. Account for the total decay width under theoretical and experimental uncertainties, $\Lambda_c(2940)^+$ is possibly the $P$-wave $\tilde\Lambda_{c2}^{ }(\frac{3}{2}^-)$ or $\tilde\Lambda_{c2}^{ }(\frac{5}{2}^-)$, it is also possibly the $D$-wave $\check\Lambda_{c3}^{2}(\frac{5}{2}^+)$ or $\check\Lambda_{c3}^{2}(\frac{7}{2}^+)$. In $\tilde\Lambda_{c2}^{ }(\frac{3}{2}^-)$ and $\tilde\Lambda_{c2}^{ }(\frac{5}{2}^-)$ assignments, the predicted total decay width ($151.9$ MeV and $130.7$ MeV) are bigger than the measured one. In $\check\Lambda_{c3}^{2}(\frac{5}{2}^+)$ and $\check\Lambda_{c3}^{2}(\frac{7}{2}^+)$ assignments, the predicted total decay width ($4.5$ MeV and $3.9$ MeV) are smaller than the measured one. However, the branching ratios $R=\Gamma(\Sigma_c(2520)^{++,0}\pi^{-,+})/\Gamma(\Sigma_c(2455)^{++,0}\pi^{-,+})$ are largely different in the two $P$-wave assignments or two $D$-wave assignments. Obviously, the measurement of the branching ratio $R=\Gamma(\Sigma_c(2520)^{++,0}\pi^{-,+})/\Gamma(\Sigma_c(2455)^{++,0}\pi^{-,+})$ is also very important for the identification of $\Lambda_c(2940)^+$.

\section{Conclusions and discussions\label{Sec: summary}}
In this work, the studies of observed $\Lambda_c(2595)^+$, $\Lambda_c(2625)^+$, $\Lambda_c(2765)^+$ (or $\Sigma_c(2765)^+$), $\Lambda_c(2860)^+$, $\Lambda_c(2880)^+$ and $\Lambda_c(2940)^+$ states are briefly reviewed. In the $^3P_0$ model, the OZI-allowed strong decay features of all these $\Lambda_c$ states are studied. Possible $1P$, $1D$ and $2S$ assignments of these observed $\Lambda_c$ states are examined. Their possible quantum numbers $J^P$ and internal structure are given based on our numerical results.

For $\Lambda_c(2595)^+$ and $\Lambda_c(2625)^+$, $\Sigma_c(2455)\pi$ are their only two-body decay modes. The branching fraction of the direct three-body decay mode $\Lambda_c^+\pi\pi$ is not large for $\Lambda_c(2595)^+$ while large for $\Lambda_c(2625)^+$, so it is impossible to learn $\Lambda_c(2625)^+$ only from the branching fraction of the two-body strong decay modes. Account for theoretical predictions of masses of excited $\Lambda_c$, $\Lambda_c(2595)^+$ and $\Lambda_c(2625)^+$ are possibly the $1P$-wave charmed baryons $\Lambda_{c1}(\frac{1}{2}^-)$ and $\Lambda_{c1}(\frac{3}{2}^-)$, respectively. The predicted decay widths are consistent with experiments.

$\Lambda_c(2765)^+$ ($\Sigma_c(2765)^+$) seems impossibly the $1P$-wave charmed baryon. It is possibly the $2S$-wave or $1D$-wave charmed baryon. The strong decay behavior of the two $2S$-wave ($\lambda$-mode excitation and $\rho$-mode excitation) baryons are similar, and it is difficult to distinguish them through their strong decays. So far, few experimental information of $\Lambda_c(2765)^+$ ($\Sigma_c(2765)^+$) has been obtained, and we have no sufficient information to learn $\Lambda_c(2765)^+$ ($\Sigma_c(2765)^+$).

$\Lambda_c(2860)^+$ seems impossibly a $2S$-wave charmed baryon, it may be the $P$-wave $\tilde\Lambda_{c2}^{ }(\frac{3}{2}^-)$ or $\tilde\Lambda_{c2}^{ }(\frac{5}{2}^-)$), it could be the $D$-wave $\check\Lambda_{c1}^{2}(\frac{1}{2}^+)$ or $\check\Lambda_{c1}^{2}(\frac{3}{2}^+)$. If $\Lambda_c(2860)^+$ has $J^P={3\over 2}^+$, it is possibly the $D$-wave $\check\Lambda_{c1}^{2}(\frac{3}{2}^+)$ with total decay $\Gamma=59.6$ MeV. In this case, the predicted branching ratio $R=\Gamma(\Sigma_c(2520)\pi)/\Gamma(\Sigma_c(2455)\pi)=2.8$. The measurement of the $R$ will be very important for the identification of $\Lambda_c(2860)^+$.

$\Lambda_c(2880)^+$ is impossibly a $1P$-wave or $2S$-wave charmed baryon, it may be a $D$-wave $\check\Lambda_{c3}^{2}(\frac{5}{2}^+)$ with $\Gamma_{total}=1.3$ MeV. The predicted branching ratio $R=\Gamma(\Sigma_c(2520)\pi)/\Gamma(\Sigma_c(2455)\pi)=0.35$, which is consistent with the measured $R=\Gamma(\Sigma_c(2520)\pi)/\Gamma(\Sigma_c(2455)\pi)=0.225\pm0.062\pm 0.0255$.

$\Lambda_c(2940)^+$ is possibly the $P$-wave $\tilde\Lambda_{c2}^{ }(\frac{3}{2}^-)$ or $\tilde\Lambda_{c2}^{ }(\frac{5}{2}^-)$, it is possibly the $D$-wave $\check\Lambda_{c3}^{2}(\frac{5}{2}^+)$ or $\check\Lambda_{c3}^{2}(\frac{7}{2}^+)$. The branching ratio $R=\Gamma(\Sigma_c(2520)\pi)/\Gamma(\Sigma_c(2455)\pi)$ are largely different in theses assignments, which could be employed to distinguish them by experiment in the future.

From Table~\ref{table2} and Table~\ref{table3}, the two light quarks couple with spin $S_\rho=0$ or spin $S_\rho=1$ in different configurations. In the assignments consistent with experiments, the two light quarks couple with spin $S_\rho=0$ in $\Lambda_c(2595)^+$ and $\Lambda_c(2625)^+$. The two light quarks couple with spin $S_\rho=1$ in $\Lambda_c(2860)^+$, $\Lambda_c(2880)^+$ and $\Lambda_c(2940)^+$. In $\Lambda_c(2765)^+$ (or $\Sigma_c(2765)^+$), $S_\rho=0$ and $S_\rho=1$ are both possible.

High excited assignments such as the $2P$-wave or $2D$-wave charmed baryons have not been examined for these $\Lambda_c$ states, and relevant calculation and analyses have not been made in the $^3P_0$ model. Other higher possible excitations assignments to these $\Lambda_c$ states may be possible. There are some uncertainties in the $^3P_0$ model. The main uncertainties result from the uncertainties of parameters $\gamma$ and $\beta$. These uncertainties may result in some large uncertainties of the numerical results. However, the predicted branching ratios depend weakly on the parameters.
\begin{acknowledgments}
 This work is supported by National Natural Science Foundation of China under the Grant No. 11475111.
\end{acknowledgments}

\begin{appendix}
\section{Flavor wave functions of baryons and mesons}
The flavor wave functions of baryons and mesons involved in our study are employed as those in Ref.~\cite{roberts}
\begin{eqnarray*}
\Lambda_c^+=\dfrac{1}{\sqrt{2}}(ud-du)c;   \Sigma_c^{++}=uuc;\\
\Sigma_c^+=\dfrac{1}{\sqrt{2}}(ud+du)c;    \Sigma_c^0=ddc;\\
p=\dfrac{1}{\sqrt{2}}(du-ud)u; \pi^+=u\bar{d};\\
n=\dfrac{1}{\sqrt{2}}(du-ud)d; \pi^-=d\bar{u};\\
\pi=\dfrac{1}{\sqrt{2}}(u\bar{u}-d\bar{d}); D^+=\bar{d}c;D^0=\bar{u}c;\\
\end{eqnarray*}
\end{appendix}

\end{document}